%% file: learning_paper.tex
\documentclass[english,12pt,draftcls,onecolumn]{IEEEtran}
\usepackage{color,graphicx,amsmath,amssymb,amsthm,epsfig,mathrsfs,cite,bm}
\include{notation}

\usepackage{amssymb}
\usepackage{float}
\usepackage{tikz}
\usetikzlibrary{trees}
\usepackage{qtree}
\usepackage{algorithm}
\usepackage{algpseudocode}
\usepackage{lipsum}
\usepackage[lofdepth,lotdepth]{subfig}

\title{Rate Prediction and Selection in LTE systems using Modified Source Encoding Techniques}

\ifCLASSOPTIONtwocolumn
\author{Saishankar K.P.$^{*}$ \hspace{2cm} Sheetal Kalyani$^{*}$ \hspace{2cm} Narendran K.$^{\dagger}$ \\
 \hspace{-2 cm}Department of Electrical Engineering$^{*}$ \hspace{1.5cm}Centre for Excellence in Wireless Technology,$^{\dagger}$\\ \hspace{-4cm} Indian Institute of Technology, Madras \hspace{3.5cm} IITM Research Park ,\\
 \hspace{-2.5cm} Chennai, India 600036\hspace{4.5cm} Chennai, India 600113\\
 \hspace{-3.5cm} \{ee09d025,skalyani\}@ee.iitm.ac.in\hspace{3.5cm} knaren@cewit.org.in
 }

\else
\author{Saishankar K.P. $^{*}$ \hspace{2cm} Sheetal Kalyani $^{*}$  \hspace{2cm} Narendran K. $^{\dagger}$  \\
 \hspace{-1.3 cm}Department of Electrical Engineering$^{*}$ \hspace{0cm} Centre for Excellence in Wireless Technology,$^{\dagger}$\\ \hspace{-4cm} Indian Institute of Technology, Madras \hspace{2.5cm} IITM Research Park ,\\
 \hspace{-2cm} Chennai, India 600036\hspace{4cm} Chennai, India 600113\\
 \hspace{-3.5cm} \{ee09d025,skalyani\}@ee.iitm.ac.in\hspace{2.5cm} knaren@cewit.org.in\\
 \hspace{8cm} 
 }
\fi
\begin{document}
 
\maketitle
\begin{abstract}
In current wireless systems, the base-Station (eNodeB) tries to serve its user-equipment (UE) at the highest possible rate that the UE can reliably decode. The eNodeB obtains this rate information as a quantized feedback from the UE at time $n$ and uses this, for rate selection till the next feedback is received at time $n+\delta$. The feedback received at $n$ can become outdated before $n+\delta$, because of a) Doppler fading, and b) Change in the set of active interferers for a UE. Therefore rate prediction becomes essential. Since, the rates belong to a discrete set, we propose a discrete sequence prediction approach, wherein, frequency trees for the discrete sequences are built using source encoding algorithms like Prediction by Partial Match (PPM). Finding the optimal depth of the frequency tree used for prediction is cast as a model order selection problem. The rate sequence complexity is analysed to provide an upper bound on model order. Information-theoretic criteria are then used to solve the model order problem. Finally, two prediction algorithms are proposed, using the PPM with optimal model order and system level simulations demonstrate the improvement in packet loss and throughput due to these algorithms. 
\end {abstract}

\section{Introduction}
4G systems, based on standards such as Long Term Evolution (LTE) offer peak data rates of upto 300 Mbps \cite{sesia2009lte} and rate adaptation through adaptive modulation has played a crucial role in facilitating this. Adaptive modulation techniques exploit the variations in the wireless channel by trying to communicate at a rate (bits per channel use), that is suited to the current channel conditions. 4G standards such as LTE supports upto 28 different rates on the downlink. The transmitter will not know the $\sinr$ at the receiver, and hence needs rate feedback from the receiver. Since we are looking at the downlink of a cellular system, the transmitter is always the Base-station/evolved NodeB (eNodeB) and the receiver is the User Equipment(UE)\footnote{In the uplink the eNodeB knows the $\sinr$ since it is the receiver.}\cite{sesia2009lte}.

The UE first measures/estimates the post-processing $\sinr$ \ie the $\sinr$ seen after receive processing such as, Minimum Mean Squared Error (MMSE) detection. Then, it calculates a rate metric which reflects the channel capacity based on standard link adaptation/abstraction techniques \cite{sesia2009lte} .  Typically, this rate metric is quantized, and LTE supports 4 bit quantization. The quantized feedback is called Channel Quality Indicator (CQI), and it is a number between 0 and 15 \cite{sesia2009lte}. The CQI feedback is done by all UEs in the the system and each UE may use different techniques for $\sinr$ measurements and rate calculations, as, these algorithms are proprietary to each receiver. The 4 bit CQI value received at the eNodeB is mapped to a 5 bit value (28 possible states) called the Modulation and Coding Scheme index (MCS). Once the CQI feedback received at time $n$ from a user $u$ is mapped to an MCS value $X_n^u$, it will be used till the next CQI feedback is received and mapped at time $n+\delta$ to $\xud$.
In this work we look at prediction of the MCS indices $\x_{n+i}$ for times $i=1,2...\delta-1$ using the discrete sequence of past values $\{\xun,\x_{n-\delta}, \x_{n-2\delta} ... \}$.
There are two reasons why prediction of MCS index is required:
\begin{enumerate}
\item The MCS available at time $n$  may have been computed from a CQI estimated by a UE at time $n-\gamma$, where $\gamma$ is the reporting delay and this shall be henceforth referred to as delayed MCS. A detailed study of the effect of CQI delay is provided in \cite{martin2007,kuhne2008throughput}.
\item The MCS available at $n$ ($\xun$) has to be used till time $n+\delta$. The channel and interference conditions can change between $n$ and $n+\delta$ leading to outdated MCS value $\xun$. Our focus in this work is on the effect of outdated MCS.
\end{enumerate}
 While the problem of delayed MCS can be addressed at the UE, the problem of outdated MCS cannot be addressed by the UE alone. This is because, if the UE were to predict and feedback the CQI meant for $n+\delta$ at $n$, the eNodeB would be left with no knowledge as to what MCS is to be used for times $n,n+1 \hdots n+\delta-1$.  Therefore, it is necessary that the eNodeB has a prediction mechanism to handle the outdated MCS problem. There are various prediction schemes \cite{cqijak,iircqi,cqiaging,marmon2007} that can be implemented at the UE which can correct for delayed CQI and complement the proposed prediction scheme used at the eNodeB.

 The MCS $\xun$ can become outdated by $n+i$, where $i<\delta$, due to the change in $\sinr$ over time because of the following reasons :
\begin{enumerate}

 \item The desired signal and interference power changes gradually over time due to Doppler effect, and the change is a function of the mobility of UEs and the scattering objects.
 \item  The active set of interfering eNodeBs for a specific UE can change over time due to the following reasons:
\begin{enumerate}

 \item The traffic patterns at the different eNodeBs may change over time, and when an eNodeB does not have enough data to send, it does not transmit over all sub-bands. For example, a user $u$ scheduled in band $i$ at time $n$ sees eNodeBs indexed as 1,5,9 as its interferers, however by $n+\delta$ a couple of eNodeBs from that set may have stopped transmitting and some other eNodeB which was inactive at $n$ may have become active at $n+\delta$ in band $i$ leading to $u$ seeing a different set of active interferers.
 \item In the case of Het-Nets, in order to reduce the interference seen by pico eNodeBs, the macro eNodeBs may not transmit on certain bands on which the pico is transmitting \cite{alrawi2012,lopez2011enhanced}. This is called sub-frame blanking and the set of active bands for an eNodeB changes dynamically when dynamic sub-frame blanking is employed \cite{alrawi2012} resulting in a change in the active set of interferers for UEs attached to neighbouring eNodeBs.  The transmission power of a macro eNodeB is 46 dBm, while that of pico is only around 23-30 dBm \cite{lopez2011enhanced}. Therefore, when the eNodeB does not transmit in some sub-frames, it ceases to be an active interferer for UEs attached to the neighboring eNodeBs and the pico power is too low for it to become a dominant interferer. 
 \end{enumerate}
 \end{enumerate}

  If the system is such that all eNodeBs transmit data always and the change is only due to Doppler, it is called a fully loaded system. On the other hand, if all eNodeBs do not transmit over all resources, it is referred to as partial loading \footnote{Note that we are looking at reuse-one LTE system where all the frequency bands are used in all eNodeBs, and in partial loading some bands may be unoccupied}. Typically, the change in $\sinr$ due to partial loading is more abrupt, leading to higher variability in MCS values.  

 There are many CQI prediction methods, proposed in \cite{cqijak,iircqi,cqiaging,marmon2007} with the objective of improving link adaptation.  In \cite{cqijak} the authors perform channel prediction using Jakes and ITU models and use it for CQI updation. In \cite{cqiaging} also channel prediction is employed to estimate the future CQI. In \cite{iircqi}, the authors treat the CQI prediction as a filtering-prediction problem, where they treat the CQI as a real number and use a linear predictor which minimizes the Mean Square Error of the CQI estimate.  It can be seen that, all the above papers, treat CQI as a continuous quantity and use filtering based prediction approaches. Furthermore, the focus is more on the effect of delayed CQI/MCS and partial loading has not been considered.

These techniques should be applied only at the UE, because, a continuous CQI viz., the actual value of $\sinr$ is available only at the UE.   At each UE, the $\sinr$-CQI mapping is done based on the receive algorithms used by it\footnote{For the same $\sinr$ different receive schemes say MRC,MMSE or ML can support different rates.}, the transmission mode and the $\sinr$ estimation itself may be different for different users \cite{holma2009lte}. This results in different receivers computing/predicting the CQI using different techniques. Since CQI is quantized, the $\sinr$ to CQI mapping is non-invertible and furthermore,  the $\sinr$ to CQI mapping employed at each UE is unknown to the eNodeB. Hence, mapping the MCS back to $\sinr$  at the eNodeB will not improve prediction accuracy. Moreover, since the eNodeB selects only a discrete rate, one can apply discrete sequence prediction, wherein, a temporal distribution of the MCS values can be built and exploited  for prediction. This technique of building the MCS distribution is practically viable only if the MCS comes from a discrete set.


We assume that the feedback is periodic with time period $\delta$ (5ms), thus the eNodeB by time $n$ will have received a sequence $\{\xun,\x_{n-\delta}...\x_{0}\}$ from the user $u$.\footnote{However, the approach proposed in this work can be modified and used even if the feedback is non-periodic or event triggered.} Our aim is to predict $\xud$ given this discrete sequence. If the joint distribution between the future and the past, i.e., $P(\x_0...\xun,\xud)$  is known, we would be able to optimally predict $\xud$ from the previously observed sequence. However, as this distribution is not known, we propose to build the joint distribution, for each user $u$. 

 We initially propose to use  algorithms from source encoding to estimate the distribution of the MCS sequence of each UE, since estimating the distribution of a source transmitting symbols, is a problem that has been studied extensively in source encoding. Certain issues in practically applying these algorithms are discussed, and appropriate modifications are proposed. In this paper, two source encoding algorithms  namely  Active Lempel Ziv (Active LeZi) and Prediction by Partial Match (PPM) \cite{gopal2007,katsaros2009prediction} are discussed. These algorithms build frequency trees and use these trees for prediction. 
 The Active LeZi algorithm converges to the optimal tree depth if one has an asymptotically long MCS sequence \cite{rissanen1978modeling,gopal2007}. However, an asymptotically long sequence may not be available in a practical system. Two major reasons for this are a)UE sleep cycles due to Discontinuous Reception(DRX) and b)the fact that the MCS sequence may not remain stationary over very long time periods. Both are discussed in detail in Section \ref{lza}. In other words, one cannot assume very long sequence lengths and, a short sequence of MCS values at the eNodeB may not be enough, for Active LeZi to converge to the optimal tree depth. Furthermore, it is also difficult to implement Active LeZi, because of a growing memory requirement even if an asymptotically long MCS sequence was available. Therefore, we propose to use PPM which uses a fixed depth frequency tree \cite{katsaros2009prediction}. 

 However, we need to know the tree depth that must be traversed for prediction using PPM. The tree depth used must capture the complexity of the sequence and at the same time the distribution built must be accurate to the depth used, given an observed sequence length.
These two requirements represent a trade-off in choosing the tree depth and the implications of this trade-off are discussed in Section \ref{mos}. 
 We propose to analyse the sequence complexity using a metric called sub-extensive information\cite{bialek2001} and use it as an upper bound on tree depth as discussed in Section \ref{sei}.  

However, as the tree depth increases, the number of parameters in the distribution required to be estimated increases. Hence, one has to optimally pick a depth that will reflect the underlying sequence complexity, and at the same time will not involve estimation of too many parameters. We propose to use classical model order estimators such as Minimum Description Length (MDL), Akaike Information Criterion (AIC) based estimators in Section \ref{genm} for finding the optimal tree depth, with the optimal model order being upper bounded by the $\kuopt$ (tree depth) given by analyzing the MCS sequence complexity. Since we have only a finite length MCS sequence available in a practical system, we focus on a finite sample corrected model order estimator to find the optimal tree depth for prediction $\kuoptt$. Note that $\kuopt$ is the optimal tree depth when the distribution is known, whereas $\kuoptt$ is the optimal tree depth when the distribution also has to be estimated.

Once the tree depth is estimated, we can build the distribution to the desired order $\kuoptt$ and use that for prediction. For the prediction step, a MAP estimator and a Bayesian Risk Minimizer are proposed for estimating $\xud$ given the MCS sequence and the estimated distribution. 

We compare the performance obtained using the proposed algorithms with a Markov predictor which uses a fixed model order across all users, the best scheme given in \cite{marmon2007} and a naive algorithm which uses the feedback without any prediction whatsoever. The work in \cite{marmon2007}, uses order statistics such as mean, median, auto-correlation etc. to perform CQI/MCS prediction at the eNodeB while, we attempt to predict MCS at the eNodeB by building a temporal distribution. 

It is possible that the CQI that has been reported may sometimes be in error as studied in \cite{aho2011cqi}. In that work, they study the effect of bias in CQI reporting and correct it using the ACK/NACK reports {\footnote{These indicate whether a packet has been received successfully or not.}} from the UE. Note that, while \cite{aho2011cqi} can correct for bias in CQI reported, it is not a prediction technique and cannot efficiently solve the problem of outdated MCS. On the other hand, while we exploit the underlying MCS sequence complexity for efficient prediction, our techniques are not designed to handle CQI error. However, our method and the method in \cite{aho2011cqi} can be easily combined in order to handle both CQI reporting error and the effect of outdated MCS.

\begin{table}[ht]
\caption{List of Symbols used}
  \centering
    \begin{tabular}{|l|l|}
        \hline
        $\xun$ & MCS index $X$ for user $u$ received at time $n$ \\ \hline
        $S^u_n$ & Sequence of MCS indices received upto time $n$ \\ \hline
        $Ipred(k)$ & Predictive information in sequence with model order $k$. \\ \hline
        $\kuopt$ & Optimal Model order as estimated using $Ipred(k)$ \\ \hline
        $\kuoptt$ & Optimal Model order when the distribution is unknown.\\
         \hline
        \hline
    \end{tabular}
\end{table}
\section {System Model}\label{symmo}
A 19 cell, 3 sectors per cell reuse-one LTE system is considered. In the system simulator, there are 19 cells and 57 sectors with wrap around, to avoid edge discontinuities \cite{codiv} and UEs are distributed uniformly in each sector.  LTE systems, use OFDMA in the physical layer where sub-carriers are grouped into sub-bands \cite{lte36211}, and users are allocated a set of sub-bands for data transmission. Each eNodeB transmits over the same set of resources, as, it is a reuse-one system. The OFDMA for the 10MHz LTE system has 1024 sub-carriers where only the 600 in the middle are used \cite{lte36211}. These 600 sub-carriers  are grouped into 50 groups of 12 sub-carriers (SCs) each and this is done over 14 OFDM symbols. So this group of 12 SCs over 14 symbols is called one Physical Resource Block(PRB) and the 14 OFDM symbols together constitute a sub-frame \cite{lte36211}. There are 50 PRBs in a sub-frame and a continuous block of 3PRBs are grouped to form a sub-band. There are 17 sub-bands in LTE for the 10MHz system \cite{lte36211}, and, scheduling and transmission is done at the sub-band level. The frame structure is provided in \cite{tran2012overview}. The set of sub-bands allocated to a user, is called a transport block and every user will be allocated one rate for the whole transport block. 

There are multiple feedback techniques in LTE and here we focus on periodic feedback, where the user combines the best five sub-bands' rates and feeds back this aggregated CQI index along with the sub-band location. This estimation of the aggregated CQI is highly UE specific \ie different UEs are manufactured by different vendors and consequently, the algorithms used may vary. At the eNodeB these CQI values are converted into MCS values. Hence, our data comprises of the MCS sequences for all the users in the system.
%
We use a full system simulator to obtain the data \ie MCS sequences for each UE used for prediction. Both, path loss exponent and shadow fading parameters are as specified in \cite{scm,lte36814} for an Urban Macro model. The channel model used in the simulator is the Generic Channel model as given in \cite{scm,lte36814}. The generic channel model is a realistic channel model for multipath channels in cellular systems. The model is such that the channel from each UE to each eNodeB is modeled using different parameters such as Angle of Arrivals and Departures of the multipath rays, distance dependent power delay profile, Line of Sight parameters and multipath profiles \cite{scm,lte36814}. Hence, different users see different delay spreads and even the same user sees different delay spreads from different eNodeBs \ie the multipath power delay profile of the channel between the UE and serving eNodeB can differ from the power delay profile between the UE and interfering eNodeBs.  This makes a simple statistical characterization of the channel for the purposes of modeling the $\sinr$ or rate extremely difficult. Even if, one were to characterize the channel, it is to be done for all the users, and the different links between eNodeBs and UEs, making it an extremely complex system to model mathematically. Note that only the strongest 8 interferers to each user, are modeled explicitly for ease of computation. The detailed simulation parameters are given in Table \ref{table:bsm} for completeness.

\ifCLASSOPTIONtwocolumn
\begin{table}
\caption{Baseline Simulation Parameters}
  \centering
    \begin{tabular}{|l|l|}
        \hline
        Deployment scenario & Urban macro-cell  scenario \\ \hline
        Base station antenna height & 25 m, above rooftop \\ \hline
        Minimum distance between & $ >= 25m $\\ UT and serving cell & \\ \hline
        Layout & 19-cell Hexagonal grid with \\&wrap around. \\ \hline
        carrier frequency & 2 GHz \\ \hline
        Inter-site distance & 500 m \\ \hline
        UT speeds of interest & 30 km/h \\ \hline
        Total eNodeB transmit& 46 dBm for 10 MHz\\ power&  \\ \hline
        Thermal noise level & -174 dBm/Hz \\ \hline
        User mobility model & Fixed and identical speed $ |v| $ \\ of all UTs,
& randomly and \\ &uniformly distributed direction \\ \hline
        Inter-site interference &Explicitly modeled \\modeling &  \\ \hline
        UT antenna gain & 0 dBi \\ \hline
        Channel Model & Urban Macro model (UMa) \\  \hline
           Network synchronization & Synchronized   \\ \hline
        Downlink transmission scheme & 1x2 Single Input Multiple Output       \\ \hline
        Downlink Scheduler & Proportional Fair with \\&full bandwidth allocation  \\ \hline
        Downlink Adaptation & sub-band Channel Quality Information \\Wideband CQI &(CQI) of best 5 bands for each user \\ 
& and for all users,at 5 ms CQI \\
&feedback periodicity, CQI delay:\\
&Ideal, CQI measurement Error: none \\
& MCS based on LTE transport formats \\ \hline
        Evaluated traffic profile & Full Loading best effort and \\ &Partial loading with \\&exponential inter-arrival time.  \\ \hline
        Simulation bandwidth & 10 + 10 MHz (FDD)   \\ \hline
        \hline
       
    \end{tabular}
    \label{table:bsm}
\end{table}

\else
\begin{table}
\caption{Baseline Simulation Parameters}
  \centering
    \begin{tabular}{|l|l|}
        \hline
        Deployment scenario & Urban macro-cell scenario \\ \hline
        Base station antenna height & 25 m, above rooftop \\ \hline
        Minimum distance between UT and serving cell & $ >= 25m $ \\ \hline
        Layout & 19-cell Hexagonal grid with wrap around. \\ \hline
        carrier frequency & 2 GHz \\ \hline
        Inter-site distance & 500 m \\ \hline
        UT speeds of interest & 30 km/h \\ \hline
        Total eNodeB transmit power & 46 dBm for 10 MHz \\ \hline
        Thermal noise level & -174 dBm/Hz \\ \hline
        User mobility model & Fixed and identical speed $ |v| $ of all UTs,\\
& randomly and uniformly distributed direction \\ \hline
        Inter-site interference modeling & Explicitly modeled \\ \hline
        UT antenna gain & 0 dBi \\ \hline
        Channel Model & Urban Macro model (UMa) \\ \hline
           Network synchronization & Synchronized   \\ \hline
        Downlink transmission scheme & 1x2 Single Input Multiple Output       \\ \hline
        Downlink Scheduler & Proportional Fair with full bandwidth allocation  \\ \hline
        Downlink Adaptation & sub-band Channel Quality Information (CQI) of best 5 bands for each user and \\ Wideband CQI 
& for all users,at 5 ms CQI feedback periodicity, CQI delay :Ideal, \\
&  CQI measurement Error: none, MCS based on LTE transport formats\\
\hline
        Evaluated traffic profile & Full Loading and Partial loading with exponential inter-arrival time.  \\ \hline
        Simulation bandwidth & 10 + 10 MHz (FDD)   \\ \hline
        \hline
        
    \end{tabular}
    \label{table:bsm}
\end{table}
\fi
The eNodeB requests MCS feedback from each user once in every $\delta$ frames (typically $\delta$=5ms), some more details are given in Table \ref{table:bsm}. Since the set of MCS values are 28, this corresponds to rates varying from 0.1523 - QPSK with code rate 0.076, to 5.5547 - 64 QAM code rate 0.93, bits per symbol \cite{sesia2009lte} seen in Table 10.1. The sequence received looks like ${X_\delta^u,X_{2\delta}^u,...X_{n}^u,X_{n+\delta}^u}..$, where the eNodeB at time instant $n+i$ ($i<\delta$) has to use a value $X_{n}^u$ which was estimated at time $n$. As discussed earlier, there are two main reasons for $X_{n+i}^u$ to vary from $X_{n}^u$ and they are a)Mobility in the system and b) The active set of interfering  eNodeBs will change.

We simulate the following traffic profiles:
\begin{itemize}
 \item A generalized traffic distribution with exponential inter-arrival rate of 50ms and packet size 3000 bytes. (partial loading)
\item A situation where all eNodeBs transmit continuously. (full loading)
\end{itemize}

To summarize, we are required to estimate, a time varying discrete value of rate, for partial and full loading. There are 57 eNodeBs with each eNodeB running scheduling algorithms independent of the other eNodeBs. These users can be scheduled over different bands, at different times, and the interfering and desired channel also changes over time.
 The above explained model is difficult to completely characterize mathematically and analyze, because, to do that we have to model the scheduler behavior under traffic, all the user-interferer channels which are not i.i.d and even time-varying traffic statistics. However, if one knows the joint temporal rate distribution  of a user, one could predict the rate from the observed sequence. Since, the sequence to be predicted is from a discrete set, we propose to use discrete sequence prediction algorithms.
%
%
%

\section {Compression Algorithms for Model Building}
In the previous sections, we explained how the MCS prediction problem for each UE could be mapped to a discrete sequence prediction problem for which a joint temporal distribution of the sequence has to be built. This problem of building a discrete distribution has been studied extensively in \cite{gopal2007,katsaros2009prediction,cleary1984data,ziv1977universal} and we propose to apply these techniques for MCS prediction with appropriate modification. We now give algorithms, which build frequency trees, and from which the discrete distribution can be estimated. 

 \subsection {Active LeZi}\label{lza}
 The Active LeZi builds a variable order Markov chain as proposed in \cite{gopal2007}. This is shown in Algorithm \ref{alz}. This algorithm uses a sliding window to update its contexts as will be explained in an example. We denote current window by $W$ its length by $W_L$ and maximum allowed window length by $W_{L_{max}}$, the dictionary by $D$ and current word as $w$.
 \begin{algorithm}[h]
 \caption{Active LeZi Algorithm} \label{alz}
\begin{algorithmic}[1]
\State $\wl=0$, $W=\emptyset$, $D=\emptyset$
\State Assign $w=\emptyset$ 
\State Append incoming character $v$ to $w$ and $W$, i.e., $W=(W,v),w=\{w,v\}$ $\wl=\wl+1$
\State If $w$ is part of $D$ do not add $w$ to $D$.
\State If $w$ is not part of dictionary add $w$ to the dictionary $D = {D,w}$.
\State $\wlm$=Maximum word length in dictionary
\State If $\wl>\wlm$ delete $W[0]$ 
\State Update frequency tree based on all contexts in the $W$.
\State Repeat from Step 2
\end{algorithmic}
\end{algorithm}
 
This algorithm generates a frequency tree for S'=22,22,22,22,22,27,27,24,24,22,24,27,24,24,22 
 as in \figref{fig:tree} and we provide an illustrative example on its working as follows:
 \begin{enumerate}[(i)]
  \item  Initialization,
  \begin{enumerate}
   \item $\wl=0$; 
   \item $W=\emptyset$; 
   \item $D=\emptyset$
   \item $w=\emptyset$  
  \end{enumerate}

    \item Getting incoming character:$v=22$
    \item From Step 3: $w=\{22\}$, $W=\{22\}$, $\wl=1$ and since ($w \notin D$) 
    \item From Step 5:$D=\{\{22\}\}$, 
    \item From Step 6 : $\wlm=1$
    \item Step 7: $\wl\ngtr\wlm$ 
    \item Step 8:
  Update tree based on $W$ as follows:\\
   \Tree[.$\emptyset$ [ .22(1)  ]] \\
   \item Step 9 - Repeating from Step 2: $w=\emptyset$, 
   \item Getting incoming character:$v=22$
   \item From Step 3:$w=\{22\}$,  $W=\{22,22\}$, $\wl=2$ and ($w \in D$ ) 
   \item Step 6: $\wlm=1$
   \item Step 7: $\wl>\wlm$: Delete $W[0]$ thus obtaining $W=\{22\}$ 
  \item Step 8: Update tree based on $W$ as follows:\\
   \Tree[.$\emptyset$ [ .22(2)  ]] \\
   \item Step 9 - Repeating from Step 2: $w=\emptyset$, 
    \item Getting incoming character:$v=22$ 
    \item Step 3: $w=\{22,22\}$, $W=\{22,22\}$, $\wl=2$ and  ($w \notin D$) 
   \item From Step 5: $D=\{\{22\},\{22,22\}\}$
   \item From Step 6: $\wlm=2$ 
   \item At Step 7: $\wl=\wlm$
   \item In Step 8: Update tree using $W$ as follows:\\
   \Tree[.$\emptyset$ [ .22(3) [.22(1) ] ]]  \\
   \item Repeat for whole sequence.
 \end{enumerate}

The full tree for the above example is shown in \figref{fig:tree}.

\ifCLASSOPTIONtwocolumn
\begin{figure}[h]
\scalebox{0.9}
{
  \Tree[.$\emptyset$ [ .22(7)   [.22(3) [ 27(1) ] ] [ .27(1) [ 27(1) ] ]  [ .24(1) [ 27(1) ] ] ] 
  	      [ .24(5)   [ .24(2) [ 22(2) ] ] [ .22(2) [ 24(1) ] ] [ .27(1) [ 24(1) ] ] ] 
  	       [ .27(3) [ .24(2) [ 24(2) ] ] [ .27(1) [ 24(1) ] ] ] ] 
  	       }
  	       \caption{Active LeZi Example Tree}
\label{fig:tree}
\end{figure}

\else
\begin{figure}[h]
  \Tree[.$\emptyset$ [ .22(7)   [.22(3) [ 27(1) ] ] [ .27(1) [ 27(1) ] ]  [ .24(1) [ 27(1) ] ] ] 
  	      [ .24(5)   [ .24(2) [ 22(2) ] ] [ .22(2) [ 24(1) ] ] [ .27(1) [ 24(1) ] ] ] 
  	       [ .27(3) [ .24(2) [ 24(2) ] ] [ .27(1) [ 24(1) ] ] ] ] 
  	       \caption{Active LeZi Example Tree}
\label{fig:tree}
\end{figure}
\fi  	       
The nodes in the tree in \figref{fig:tree} gives information about the MCS index and the number of times it has occurred in a certain MCS sub-sequence. For example, in \figref{fig:tree} if one looks at the left most node in the bottom most generation a value 27(1) is seen. This implies that the subsequence \{22,22\} has been followed by a \{27\} i.e., \{22,22,27\} has occurred once and from the parent of that node \{22,22\}, has occurred thrice , and \{22\} itself has occurred seven times.

 However, this algorithm suffers from certain implementation difficulties. The maximal word length in this algorithm grows with sequence length, thereby, requiring an ever-increasing memory to store the words and frequency trees. Since the channel correlations are typically of the order of only a few milliseconds, the correlations in the MCS sequences does not extend much in time and, it is unnecessary to learn very long contexts to predict.

Furthermore, this predictor converges to the optimal model order only asymptotically \cite{rissanen1978modeling}. However, due to the effect of UE sleep cycle, we would never see an asymptotically long sequence to learn the data \cite{sesia2009lte}. In order to save battery, when the user is idle it stops measuring/sensing the channel and hence there is no feedback during this time. There are two types of sleep cycles viz. short DRX or long DRX. First, the UE senses the control channel, to know, if there is any data to be received and if there is no data to be received it goes into a short sleep cycle, where the UE does not sense the channel or feedback MCS. Then, it again senses the channel at the end of the short DRX and if there is still no data it goes for another short DRX and after $N$ such short DRX, if there is no data the UE goes into long DRX. The length and duration of short and long DRX and $N$ are configurable, and are configured according to traffic type that the UE is receiving. Furthermore, the assumption of stationarity may not hold over very long time periods or sequence lengths. Hence in a practical system, one has to assume that the sequence length is limited.

Since Active LeZi requires a high amount of memory and also an asymptotically long sequence, both of which are not practical, we propose to use the PPM method of a fixed tree depth with appropriate modifications.
%
\subsection{Prediction by Partial Match}\label{ppm}
Most online predictors are based on the short memory principle, in which the recent past is more important for prediction i.e. prediction is done by observing the previous $k$ symbols. Here, we plan to build a frequency tree of fixed depth $k^{max}$ which may depend on the sequence length available. The PPM uses the Active LeZi algorithm with the $\wlm$ fixed to some $k^{max}$. Now using PPM, with fixed tree depth $\km$, one can compute all models of the form $P(\xun|\x_{n-\delta} ... \x_{n-k\delta})$ with $k={1,\hdots,\km-1}$. Note that, while one can build the tree upto depth $\km$, the depth used for prediction can be different. This depth used for prediction will depend on the sequence complexity and the number of parameters one needs to estimate to learn the distribution (details given in Section \ref{mos}).  The example tree given in \figref{fig:tree} has  $\km=3$ and from this tree, the models $P(\xun|\x_{n-\delta})$ and $P(\xun|\x_{n-\delta},\x_{n-2\delta})$ can be computed and either of them can be used for prediction.
\subsection{Estimation of $P(\xun|\x_{n-\delta} ... \x_{n-k\delta})$ using the Frequency Trees}
 Using the techniques presented above, Markov models upto order $\km-1$ can be built. In order to use a $k$th order model to predict, each state needs to be assigned a probability of occurrence, given the model and previous $k$ states. This has to be done using the models of order $1$ to $k$ which are recursively built. This recursion is because even if a $k$th order model returns the probability of a particular state as zero, there might be a lower order context in which the state could have occurred. For instance, if one looks at the example sub-sequence given in Section \ref{lza} and its corresponding tree in \figref{fig:tree}, from the second order model alone, the next value being {22} is zero because, {24,22} has never been followed by a {22}. However, if one looks at the first order model, {22} has succeeded a {22}, 3 out of 7 times. Therefore, the information upto depth $k+1$ must be blended to give the probability of occurrence of a state under model order $k$. Typical blending methods are given in \cite{gopal2007,begleiter2004prediction}. Given the frequencies of all contexts and given that the previous $k-1$ alphabets were $\x_{n}\hdots\x_{n-k+2}$ then the probability that the next state is $\xud=t_i$ is given by a recursive computation.
 \ifCLASSOPTIONtwocolumn
\begin{align}
&P_0(\xud=t_i)=\frac{\sum_{i=1}^{n} \one(\x_{i}=t_i)}{n} \label{eq:p1}\\
 &P_k(\xud=t_i)=\nonumber\\& P(\xud=t_i|\xun,..,\x_{n-(k-1)\delta}=t_{j_1}..t_{j_k}) \nonumber \\ &= \frac{\sum_{i=1}^{n}\one(\x_{(i+k)\delta}, ...\x_{i\delta}=t_j..t_{j_k})}{\sum_{i=1}^{n}\one(\x_{(i+(k-1))\delta}, .. \x_{i\delta}=t_{j_1}..t_{j_k})} \nonumber \\  &+ \left(1-\frac{\sum_{t_j}\sum_{i=1}^{n}\one(\x_{(i+k)\delta}..\x_{i\delta}=t_j..t_{j_k})}{\sum_{i=1}^{n}\one(\x_{(i+(k-1))\delta}..\x_{i\delta}=t_{j_1}..t_{j_k})}\right)\cdot \nonumber \\&P_{k-1}(\xud=t_i) \label{eq:p2}
\end{align} 
 \else
\begin{align}
P_0(\xud=t_i)&=\frac{\sum_{i=1}^{n} \one(\x_{i}=t_i)}{n} \label{eq:p1}\\
 P_k(\xud=t_i)&= P(\xud=t_i|\xun,..,\x_{n-(k-1)\delta}=t_{j_1}..t_{j_k}) \nonumber \\ &= \frac{\sum_{i=1}^{n}\one(\x_{(i+k)\delta}, ...\x_{i\delta}=t_j..t_{j_k})}{\sum_{i=1}^{n}\one(\x_{(i+(k-1))\delta}, .. \x_{i\delta}=t_{j_1}..t_{j_k})} \nonumber \\  &+P_{k-1}(\xud=t_i)\cdot  \left(1-\frac{\sum_{t_j}\sum_{i=1}^{n}\one(\x_{(i+k)\delta}..\x_{i\delta}=t_j..t_{j_k})}{\sum_{i=1}^{n}\one(\x_{(i+(k-1))\delta}..\x_{i\delta}=t_{j_1}..t_{j_k})}\right)\label{eq:p2}
\end{align} 
\fi
where $\one$ is the indicator function, indicating the occurence of the event, and,  $\sum_{i=1}^{n}\one(\x_{(i+k)\delta}, ...\x_{i\delta}=t_j..t_{j_k})$ is the frequency of occurrence of the sequence $\{t_{j_k},t_{j_{k-1}}...t_{j_1},t_{j} \}$ where $n$ is the sequence length that has been observed.
As an example let us use the tree given in Section \ref{lza} to compute the probability that the next value of the sequence S' is 24. \\
The last seen values are 24,22 . The number of times 24,22,24 has occurred given 24,22 has occurred is 1 and the number of times that 24,22 has occurred is 2. The number of times 24,22 has occurred with no future stored context is also 1 which is the second term in  \eqref{eq:p2}. This is the probability by which the lower order model is weighed.
 Therefore
 $P(24|24,22)=\frac{1}{2}+(1-\frac{1}{2})P(24|22)$ and $P(24|22)=\frac{1}{7}$. Thus, the probability that $P(24|24,22)=\frac{1}{2}+(1-\frac{1}{2})\frac{1}{7}=\frac{4}{7}$

To summarize this section, we saw three algorithms which built frequency trees and a method to evaluate the $k$th order probability. It can be seen that, to build a $k^u$th order model for user $u$ viz. $P(X_n^u|X_{n-k^u+1}^u,X_{n-k^u+2}^u, .. X^u_{n-1}) $, one must use the data upto depth $k^u+1$ from the tree. Our next problem is finding out, the optimal $k^u$ that can used for prediction for each user $u$ (different users can have different values of $\ku$) called the {\bf model order} selection problem. In the next section, we shall discuss the model order problem in detail and propose methods to find the optimal order.
\section{Model Order Selection based on Sequence Complexity and AIC} \label{mos}
The algorithms which built frequency trees and evaluated probabilities using them were discussed in detail in the previous section, and now we want to find out the depth of the tree upto which one has to traverse, to obtain a 'reasonable model'. 

{ A model used for prediction must satisfy two properties:
\begin{itemize}
 \item The model used must capture the complexity of the sequence.
 \item The frequency tree built, must be 'reasonably' accurate to the required depth, given an observed sequence length.
\end{itemize}
The first property is intrinsic to the sequence, i.e. a sequence comes from a particular distribution $P(X_{(N-k+1)\delta}^u,X_{(N-k+2)\delta}^u,...X_{(N-k+i)\delta}^u...X_{N\delta}^u)$ such that given the previous $k^u-1$ values, any knowledge of values further in the past does not improve the prediction accuracy. The second property arises due to the fact that the distribution is being estimated, and with increasing $k^u$, the number of parameters to be estimated increase and to estimate a large number of parameters a correspondingly large sequence must be observed. In other words, if the model that best  fits a given sequence is $k^*$, it could be that the number of parameters to be estimated for building a $k^*$ model might be so large that estimating the required parameters accurately from a fixed length MCS sequence may not be possible. Hence, the optimal model order is that, which achieves the right balance, in the trade-off between, finding a model which is complex enough to capture the sequence complexity, but not so complex that it requires a large number of parameters to be estimated.} These two properties are explained in detail in the next subsections. For the sake of notational simplicity, we henceforth drop $\delta$ from the subscript \ie $\x_{i\delta}=\x_i$
\subsection{Sub-Extensive Information as a metric for Sequence Complexity}\label{sei}
We first focus on a metric which characterizes the underlying complexity/ learnability/ predictability of a sequence called sub-extensive information \cite{bialek2001}. We had mentioned earlier that, sequence prediction is similar to source encoding and hence, it is only natural that, we study the model order through complexity and entropy of the sequences. The absolute entropy of a sequence increases with volume per se because complexity scales with volume \cite{cover2012elements}. Since, sequence prediction involves predicting the future, having observed the past, one is more interested in the mutual information between the past and the future than the absolute entropy. This mutual information is also called sub-extensive information or predictive information in sequence prediction literature in physics \cite{bialek2001}. The total information/entropy in a sequence is a sum of extensive and sub-extensive information components. The total entropy at time $n$ is given by: 

\ifCLASSOPTIONtwocolumn
\begin{align}
 H(X_{total})&=H(\x_1,\x_2,\x_3,...,\xun)\nonumber\\
 &=H(\xun|\x_{n-1}..X_1^u) \nonumber\\ &+ H(\x_1,\x_2,..,\x_{n-1}) \label{eq:ent}
\end{align}

\else

\begin{align}
 H(X_{total})&=H(\x_1,\x_2,\x_3,...,\xun)\\
 &=H(\xun|\x_{n-1}..X_1^u) + H(\x_1,\x_2,\x_3,...,\x_{n-1}) \label{eq:ent}
\end{align}
\fi
The first term on the RHS of \eqref{eq:ent} is the sub-extensive component and the second term is the extensive component of entropy. It can be seen that, as $n\longrightarrow \infty$ the total entropy and the extensive component will tend to infinity linearly with $n$, while the sub-extensive component will grow at a less than linear rate
The average sub-extensive/mutual information is given by:
\ifCLASSOPTIONtwocolumn
\begin{align}
  &I(\x_n,(\x_1,\x_2,\x_3,...,\x_{n-1}))\nonumber \\ &=\left\langle log_2\left(\frac{P(\x_n|(\x_1,\x_2,\x_3,...,\x_{n-1}))}{P(\x_n)} \right)\right\rangle \label{eq:ipr}
\end{align}
\else
\begin{align}
  I(\x_n,(\x_1,\x_2,\x_3,...,\x_{n-1}))=\left\langle log_2\left(\frac{P(\x_n|(\x_1,\x_2,\x_3,...,\x_{n-1}))}{P(\x_n)} \right)\right\rangle \label{eq:ipr}
\end{align}
\fi
where, $\left\langle \right\rangle$ denotes expectation over the joint distribution, $P(X_1 .. X_n)$.
Another way of writing this is:
\ifCLASSOPTIONtwocolumn
\begin{align}
 &I(\x_n,(\x_1,\x_2,\x_3,...,\x_{n-1}))=\nonumber\\&H(\x_n)+ H(\x_1,\x_2,\x_3,...,\x_{n-1})\nonumber \\&-H(\x_1,\x_2,\x_3,...,\x_n) \\
&I(\x_n,(\x_1,\x_2,\x_3,...,\x_{n-1}))=H(\x_n)\nonumber \\&-H(\x_n|\x_1,\x_2,\x_3,...,\x_{n-1}) \label{eq:bou}
\end{align}
\else
\begin{align}
 I(\x_n,(\x_1,\x_2,\x_3,...,\x_{n-1}))&=H(\x_n)+ H(\x_1,\x_2,\x_3,...,\x_{n-1})\nonumber
\\&-H(\x_1,\x_2,\x_3,...,\x_n) \\
I(\x_n,(\x_1,\x_2,\x_3,...,\x_{n-1}))&=H(\x_n)-H(\x_n|\x_1,\x_2,\x_3,...,\x_{n-1}) \label{eq:bou}
\end{align}
\fi
Calculating the sub-extensive part of information requires the knowledge of joint probability distributions. This sub-extensive component of information, is also called predictive information and is denoted as:
\begin{align}
 \Ipred(T,T')=\left\langle log_2\left(\frac{P(\x_{future}|\x_{past})}{P(\x_{future})} \right)\right\rangle \label{eq:iprid}
\end{align}
where $T$ is the time for which the sequence has been observed in the past and $T'$ is the future time for which the sequence is to be predicted. Computing the $\Ipred(T,T')$ as in  \eqref{eq:iprid} requires the knowledge of the joint distribution of the entire sequence. However, in practical systems one may not have the complete joint distribution of $\{\x_n,\x_{n-1}..\x_{1} \} $ and due to memory constraints, it will be possible to estimate and use only the joint distribution of $\{ \x_n,\x_{n-1}..\x_{n-k} \}$ . 

In our problem the focus is on finding the best $k^u$-th order Markov model for each user $u$, to use in PPM for prediction, and the predictive information in a sequence while using a model of order $k$ is denoted by $\Ipred(k)$. The value of $k$ can be varied from $1$ to $K$ and $\Ipred(k)$ can be obtained as follows:
\begin{align}
 \Ipred(k)&=\left\langle log_2\left(\frac{P(\x_n|(\x_{n-1}..\x_{n-k}))}{P(\x_n)} \right)\right\rangle \label{eq:iprk} \\
 &= H(\x_n)-H(\x_n|(\x_{n-1}..\x_{n-k}))) \label{eq:ipen}
 \end{align}
Since, the sequence that we are studying is a sequence of MCS indices and the dependence on the past is of a decreasing nature i.e. $\x_n$ to `depends more' on $\x_{n-k}$ than $\x_{n-(k+1)}$, where $k>0$, we can expect $\Ipred(k)$ as a function of $k$ to grow at a rate slower than linear increase. $\Ipred(k)$ will be monotone non-decreasing in $k$ because the mutual information is not going to decrease as the number of observations increase. As, the number of observations used for prediction increases i.e. between using $k$ past values and using one more value in the farther past can only either increase, or retain the existing information about the future. For $\Ipred(k)$ to have a linear growth rate it would require $\x_n$ to `depend equally' on $\x_{n-l}$ and $\x_{n-(l+1)}$ which will not happen, because, both desired and interference channel correlations decrease over time and the MCS sequence depends on both. Sub-linear rate of increase can mean either a rate of increase of $O(k^\alpha)$ where $\alpha<1$ or a rate of increase of $O(log(k))$. Another possibility is that the sub-extensive information is constant despite increasing the number of observations. This can happen when the underlying process is a simple Markov process. While trying to predict a simple Markov process it is enough that we observe the immediate past, \ie $\x_{n-1}$ \cite{feller2008introduction,papoulis1991probabilities,cover2012elements}.
\subsubsection{Sub-Linear $O(k^\alpha)$ rate of increase} \label{sec:sul}
The generalized form $\Ipred(k)$, is \cite{bialek2001}:
\begin{align}
\Ipred(k)&=C_0+C_1k^\alpha  \\
L(k)&=\Ipred(k)-\Ipred(k-1) \\
L(k)& \approx \frac{\partial I_{pred}(k)}{\partial k} = \alpha C_1k^{\alpha-1} \label{eq:subl}
\end{align}
where $0<\alpha<1$.
 The term $L(k)$ is called the learning curve, and is a metric which gives the rate at which the predictive information increases when the model order is increased, and this is a decreasing function in $k$ from  \eqref{eq:subl}. This implies that increasing k more and more gives only diminishing returns in prediction performance. A sub-linear rate of increase as shown in \eqref{eq:subl}, implies that the number of parameters to be learnt for predicting the sequence is infinite \cite{bialek2001}. In the problem studied here, since the sequence to be predicted itself is discrete, only finite parameters will be required to be estimated and hence, sub-linear increase will never be seen.
\subsubsection{Logarithmic $O(log(k))$ rate of increase} \label{sec:log}
The generalized form $\Ipred(k)$, is \cite{bialek2001}:
\begin{align}
\Ipred(k)&=C_0+C_1log(k)  \\
L(k)&=\Ipred(k)-\Ipred(k-1) \\
L(k)&\approx \frac{\partial I_{pred}(k)}{\partial k} =  \frac{C_1}{k} \label{eq:logl}
\end{align}
A log-rate of increase in predictive information implies that the number of parameters to be estimated is finite \cite{bialek2001}. The MCS sequences can at most have only a logarithmic rate of increase, since in predicting discrete sequences, it is required to predict only a finite number of parameters to characterize these sequences.

\ifCLASSOPTIONtwocolumn
\begin{figure}
\begin{center}
 \includegraphics[width=\columnwidth]{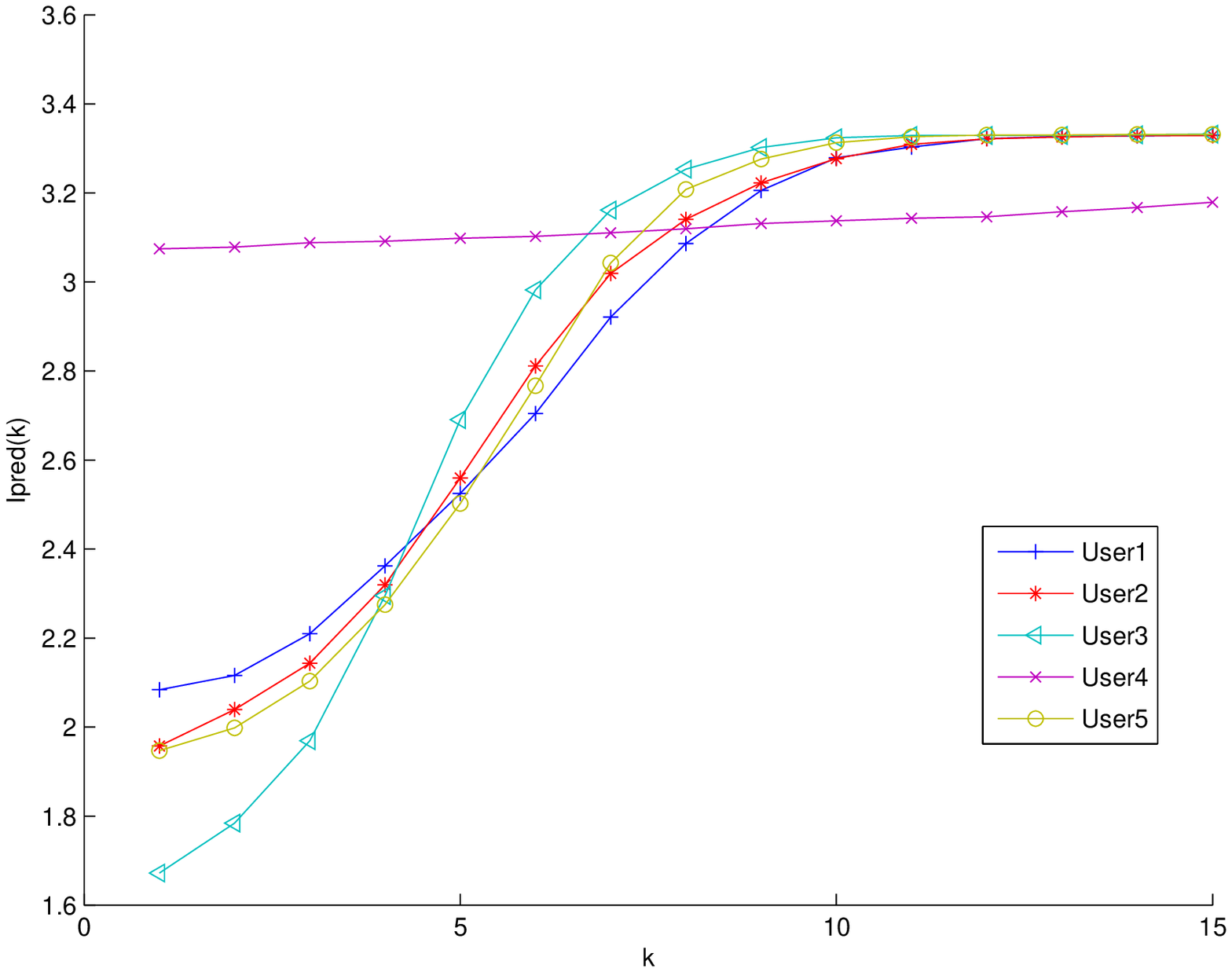}
\caption{Plot of $\Ipred(k)$ as a function of $k$}
\label{fig:ipred}
\end{center}
\end{figure}
\else
 \begin{figure}
\begin{center}
 \includegraphics[width=0.5\columnwidth]{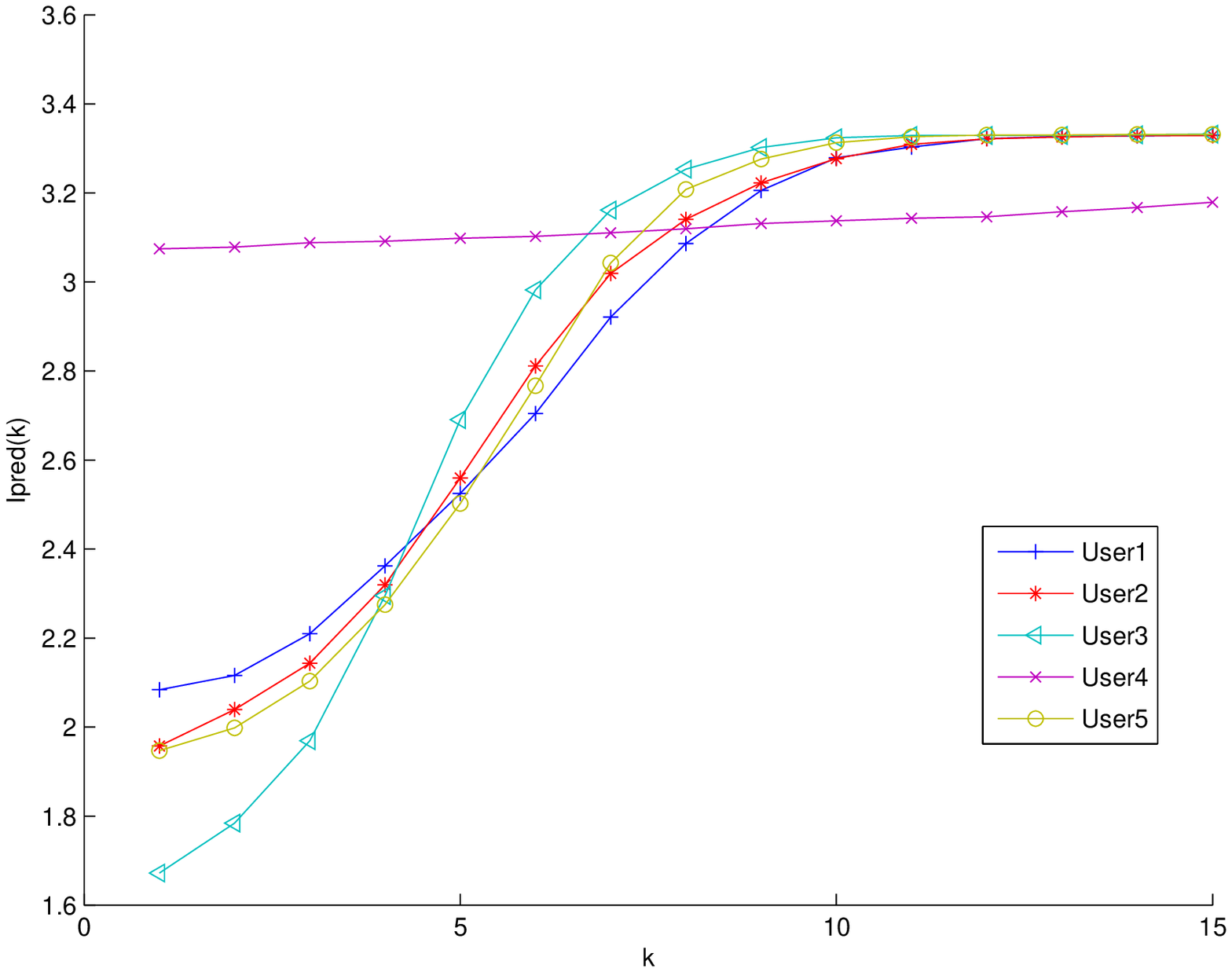}
\caption{Plot of $\Ipred(k)$ as a function of $k$}
\label{fig:ipred}
\end{center}
\end{figure}
\fi
We now compute the $\Ipred(k)$ for all the users and a few users' behaviour is captured in \figref{fig:ipred}. This computation is performed by empirically averaging the term $log_2\left(\frac{P(\x_n|(\x_{n-1}..\x_{n-k}))}{P(\x_n)} \right)$ as shown in  \eqref{eq:iprk}. The results seem to show a logarithmic behaviour, but, instead of continuously diverging the $\Ipred(k)$ saturates at a constant value.  This can be understood better by looking at  \eqref{eq:ipen}. The value of $H(\x_n|\x_1,\x_2,\x_3,...,\x_{n-k})$ is bounded from above by $H(\x_n)$ and below by $0$ and $H(\x_n)$ itself is bounded above by $log(p)$ where $p$ is the number of possible states that $\x_n$ can take \cite{feller2008introduction}. This is expressed concisely as:
\begin{align}
 0 \leq H(\x_n|\x_{n-1}..\x_{n-k}) \leq H(\x_n) \leq log(p) \label{eq:bo}
\end{align}
From \eqref{eq:ipen} and \eqref{eq:bo} it is apparent that:
\begin{align}
 0 \leq \Ipred(k) \leq log(p)
\end{align}
It can be argued that, by picking a value of $k$ for which $\Ipred(k)$ achieves its maximum possible value would give us an optimal prediction performance. However, the distribution is unknown to us and, as $k$ increases, the number of parameters needed to estimate the unknown distribution also increase and hence, the $\Ipred(k)$ that has been computed may not be accurate given the sequence of limited length. For example, in \figref{fig:ipred}, despite the sequence of User 4 having only a slowly increasing value of $\Ipred(k)$ when compared to the other users, it is the sequence that has the best prediction performance. This is because, User 4 requires only a simple Markov model to predict its sequence, and it is significantly easier to estimate the parameters of a simple Markov model as compared to estimating a model of order $4$. However, one can use the sub-extensive information to find out the maximum possible model order where the gains are substantial \ie the maximum model order $\kuopt$ can be found out as:
\begin{align}
 \kuopt=max(k):L(k)>\epsilon \label{eq:lk}
\end{align}
where $\epsilon$ is chosen such that, the gains obtained in increasing the model order beyond $\kuopt$ is not significant. For instance the User 4, will have $\kuopt=2$. 

The $\kuopt$, as calculated here is optimum if the true distribution is known to us a priori. However, we have to estimate/learn the distribution and, as $\kuopt$ of a given user increases, the number of parameters required to be estimated in order to learn the distribution increase. The effect of estimating a large number of parameters, on model order is discussed in the next section. We use the $\kuopt$ obtained in the current section as an upper bound on the optimal model order when the distribution is to be estimated.
\subsection{Optimal Model Order when the distribution is to be estimated} \label{genm}
Now, we are to fit a model order given the sequence and the distribution estimated from the sequence. The model order fitting problem is approached as a hypothesis testing problem, where $\mathcal{H}_i$ is the hypothesis that the $i$th order Markov chain best fits the sequence. Then, the optimal value of $i$ denoted by $\kuoptt$ can be found out by maximizing information theoretic criteria such as Minimum Description Length (MDL) or Akaike Information Criteria (AIC) \cite{bozdogan1987model,rissanen1978modeling,merhav1989estimation}. In these methods, the usual technique followed is to maximize the likelihood of the observations given the hypothesis, with a penalty on the number of parameters to be estimated. In the problem considered, the observation is the MCS sequence $S^u_n=\{ ..\x_m,\x_{m+\delta} ... \xun \}$ observed for each user $u$ and the number of parameters is the number of distribution parameters to be estimated. We are interested in building a discrete probability distribution of $i$ length sequences. The parameters required for building such a distribution is denoted by $\thet$ where $i$ is the model order and the cardinality of $\thet$ is $\niu$, which is the number of parameters to be estimated. $\thet$ is the $i$th order distrbution itself. For example, in our scheme, to estimate the distribution $P(\xud)$, since there are $28$ MCS values one needs to estimate $28-1$ probabilities. To estimate $P(\xud|\xun)$, one must estimate a transition probability matrix of size $(28-1)28$. By induction, this logic can be extended to an $i$th order model and the number of parameters would be $(28-1)28^{i-1}$. To generalize, if one had to estimate a $k$th order Markov Model for an $m$ state process, then $(m-1)m^{k-1}$ parameters would have to be estimated.
We use the value obtained from our $\Ipred(k)$ calculations to determine the maximum possible model order $\kuopt$ for user $u$ and use it as an upper bound on the model order to be determined.

 The model order problem can be set-up as a multiple hypothesis testing problem as follows:
\begin{itemize}
 \item $\H_1$ : Hypothesis that $\kuoptt=1$
 \item $\H_2$ : Hypothesis that $\kuoptt=2$ 
 \\ \vdots
 \item $\H_{\kuopt}$ : Hypothesis that $\kuoptt=\kuopt$
\end{itemize}

 In usual hypothesis testing problems, the likelihood function of the observations given the hypothesis is found out and the hypothesis that maximizes the likelihood function is taken to be the true hypothesis. However, when the hypotheses are models of an increasing order, this technique fails because, the lower order models are always nested within the higher order models \cite{kay1998fundamentals}. Since, we know that the error in estimating the parameters of a higher order model will also impact the performance of a system, we look at a cost function which picks a model that provides a trade-off between maximizing the likelihood and minimizing the error variance of the parameters to be estimated.

 Therefore, we propose to use the Generalized Maximum Likelihood Estimator (GMLE) in \cite{kay1998fundamentals} which tries to maximize the following cost function:
 \begin{align}
 \xi_i^u&=ln (P({\bf S^u_n};\hat{\thet}|\mathcal{H}_i))-\frac{1}{2}ln(det({\bf I}({\thet}))), \: \: \: \: 1 \leq i \leq \kuopt \label{eq:genml}, 
 \end{align}
 where the first term in \eqref{eq:genml} is the log-likelihood function and the second term is the penalty due to errors in model where ${\bf I}({\thet})$ is the Fisher information matrix of $\thet$, and its inverse is the lower bound on the error covariance matrix in estimating $\thet$, where $\thet$ is a vector of distribution parameters which are to be estimated and its cardinality is $\niu$. This set of estimates is denoted by $\hat{\thet}$ where $\hat{\thet}$ is the ML estimate of $\thet$. 


When $i$ increases, the first term in  \eqref{eq:genml} \ie the log-likelihood function increases while in the second term, because the number of parameters to be estimated increases, the $det({\bf I}({\thet}))$ increases. Therefore, maximizing the above equation with respect to $i$ ensures that, a model is choosen by optimally trading off, model likelihood with model parameter estimation error.
\begin{align}
 \kuoptt&=\underset{i} \arg \max (\xi_i^u).
\end{align}

However, to implement the above solution one must know ${\bf I}({\thet})$. That involves knowing the probability distribution function a priori. However, in our case the parameters to be estimated are the probabilities themselves. Therefore, instead of trying to estimate ${\bf I}({\thet})$, the determinant $det({\bf I}({\thet}))$ can be approximated as $cN^{\niu}$ as in \cite{kay1998fundamentals}. This is equivalent to MDL as in \cite{rissanen1978modeling} and \cite{kay1998fundamentals}.

\begin{align}
 MDL_i^u&=-ln (P({\bf S^u_n};\hat{\thet}|\mathcal{H}_i))+\frac{\niu}{2}ln(N), \: \: \: \: 1 \leq i \leq \kuopt\label{eq:mdl}.
 \end{align} 

The optimal model is obtained as:
\begin{align}
  \kuoptt&=\underset{i} \arg \min (MDL_i^u).
\end{align}

Another option is to use the AIC which is given follows:

\begin{align}
  AIC_i^u=-2ln (P({\bf S^u_n};\hat{\thet}|\mathcal{H}_i))+2\niu, \: \: \: \: 1 \leq i \leq \kuopt\label{eq:aic}.
\end{align}
Here again the optimal model order is obtained as:
\begin{align}
   \kuoptt=\underset{i}\arg \min (AIC_i^u).
\end{align}

AIC is an efficient model order estimator, while, MDL is a consistent estimator \cite{claeskens2008model}. However, both AIC and MDL assume that the number of observations is asymptotically large \ie $n \gg \niu$ \cite{claeskens2008model,hurvich1989regression}.

However, we have only finite length data sequences, and $\niu$ grows nearly exponentially in $i$. Therefore we use a sample corrected AIC \ie $AIC_C$ which is given as follows \cite{claeskens2008model,hurvich1989regression} :
\ifCLASSOPTIONtwocolumn
\begin{align}
   AIC_{Ci}^u&=-2ln (P({\bf S^u_n};\hat{\thet}|\mathcal{H}_i))+2\niu\nonumber\\&+\frac{2\niu(\niu-1)}{N-\niu-1} \: \: \: \: 1 \leq i \leq \kuopt \label{eq:aicc} \\
   \kuoptt&=\underset{i}{\operatorname{argmin}} (AIC_{Ci}^u)
\end{align}
\else
\begin{align}
   AIC_{Ci}^u&=-2ln (P({\bf S^u_n};\hat{\thet}|\mathcal{H}_i))+2\niu+\frac{2\niu(\niu-1)}{N-\niu-1}, \: \: \: 1 \leq i \leq \kuopt \label{eq:aicc}, \\
\kuoptt&=\underset{i} \arg \min (AIC_{Ci}^u).
\end{align}
\fi
The sample corrected AIC is derived in detailed in \cite{cavanaugh1997unifying}. It can be seen that the sample corrected AIC tends to the asymptotic AIC as $N \rightarrow \infty $. Also, this criterion ensures that, one does not pick a higher order model initially when the sequence length is small.

Summarizing, we have proposed usage of finite sample model order determination methods to find the best model to be used in our PPM algorithm for predicting the sequence for a given user $u$. This is to be done for all user sequences as different sequences will have different complexity. In a system like LTE there are $28$ MCS values that can occur. Therefore, to build a model of order $i$, it seems that one has to estimate nearly $28^i$ probabilities for all possible sequences. However, a user $u$ will not see all the MCS indices, in the short time frame, that we look at for sequence prediction. For instance, a user that sees MCS index $1$ corresponding to rate $0.15$ cannot see MCS 28 corresponding to rate $5.55$ within a time frame of few seconds or even between two sleep cycles. It may be that, a user sees only $m_u$ MCS indices. The value of $m_u$ is estimated from the frequency tree. For instance, consider the tree given in Section \ref{lza}. Since the only values observed in the sequence S for building the tree was {22,24,27} the value of $m_u$ will be estimated as 3. Thus for a given user $u$, finally the model order is estimated by minimizing the  cost function given below.
\ifCLASSOPTIONtwocolumn
\begin{align}
   AIC_{C i}^u&=-2ln (P({\bf S^u_n};\hat{\thet}|\mathcal{H}_i))+2(m_u-1)(m_u)^{i-1}+ \nonumber \\ &\frac{2(m_u-1)(m_u)^{i-1}((m_u-1)(m_u)^{i-1}-1)}{N-(m_u-1)(m_u)^{i-1}-1} \nonumber \\ & 1 \leq i \leq \kuopt \label{eq:aiccmod}
\end{align}

\else
\begin{align}
   AIC_C(i^u)&=-2ln (P({\bf S^u_n};\hat{\thet}|\mathcal{H}_i))+2(m_u-1)(m_u)^{i-1}+ \nonumber \\ &\frac{2(m_u-1)(m_u)^{i-1}((m_u-1)(m_u)^{i-1}-1)}{N-(m_u-1)(m_u)^{i-1}-1} \: \: \: \:  1 \leq i \leq \kuopt \label{eq:aiccmod},
\end{align}
\fi
and the optimal model order is given by:
\begin{align}
  \kuoptt=\underset{i} \arg \min AIC_C(i^u) \label{eq:aim}.
\end{align}
We have observed that when $\kuopt$ is 4, $\kuoptt$ can vary from 1 to 4.
%
%

\section{Prediction Algorithms using the Estimated Distribution}

The model order obtained in the previous sections can be used in the PPM algorithm to fix the tree depth for prediction and the probabilities $P(\xud|\S^u_n)$ can be calculated using the \eqref{eq:p1}and \eqref{eq:p2}. We now propose two prediction algorithms.
\subsection{MAP Estimator}\label{map}
The Maximum A Posteriori (MAP) estimator is an estimator that maximizes the a posteriori probability of an event given the observations \ie it picks that value which is the most likely given that the past has been observed. The MAP estimator for MCS index given the sequence observed is as follows:
\begin{align}
 \xudh=\underset{i} \arg \max P(\x_{n+1}=i|\xun .. \x_{n-\kuoptt}) \label{eq:map}
\end{align}
where $\x_{n+1}$ is the next state which we want to predict and $i$ s  are the possible values taken by the MCS.
This technique will result in maximum prediction accuracy. However, since it is optimized only for prediction accuracy, it treats all errors equally \ie estimating a rate higher than the the true rate is same as estimating a lower rate. However, in the rate prediction problem, if the predicted rate is lower than the true rate, the transmission at the predicted rate will still be a success at the cost of a loss in efficiency whereas, if the predicted rate is higher it will result in a packet loss. The MAP estimator is oblivious to this effect and therefore, will not be throughput optimal despite its prediction optimality. For instance, given a sequence $S$, if there are $3$ rates $r_1<r_2<r_3$ which are possible future candidates with probabilities $P(r_1)=0.3,P(r_2)=0.3,P(r_3)=0.4$, then the MAP estimator will pick $r_3$. Now, based on the observed data, there is approximately $60\%$ probability that $r_3$ was a wrong prediction resulting in packet loss. If the rates $r_1,r_2$ comparable to $r_3$, one could have chosen the lower rates $r_1$ or $r_2$, thus decreasing the risk of packet loss. The next section proposes a method of predicting rate given the issues of packet loss and throughput efficiency.
\subsection{Bayesian Risk based Estimator}\label{brm}
In this technique, a cost is assigned to the event of predicting a state and the state which has the minimum cost is picked. There are numerous ways of assigning the costs, and the cost assignment is done in order to enable the picking of the highest possible rate without resulting in failed transmission. The cost assignment used is as follows:
\begin{itemize}
\item If predicted rate is greater than the true rate then we lose the true rate and this is taken to be the cost of choosing the predicted rate.
\item If predicted rate is less than the true rate the difference in rate is the cost of using the predicted rate.
\end{itemize}
The expected cost of transmitting at a rate $r_{j}$ denoted by $C_j$ is given by:
\begin{align*}
C_j=\sum_{i=1}^{p} C_{ij} P(\x_{n+1}=i|\xun .. \x_{n-\kuoptt}) 
\end{align*}
where 
\begin{align}
C_{ij}=
\begin{cases}
 r_i, \:\: r_i<r_j \\
 r_i-r_j, \:\: r_i\geq r_j
\end{cases}
\end{align}
 Here $P(\x_{n+1}=i|\xun .. \x_{n-\kuoptt})$ is the probability of the system being in state $i$ given that the sequence $\xun .. \x_{n-\kuoptt}$ was observed, calculated using  \eqref{eq:p1},\eqref{eq:p2}. The predicted value of $\x_{n+1}$ is given by minimizing the expected cost $C_j$.
\begin{align}
 \xudh=\underset{j} \arg \min C_j  \label{eq:cost}
\end{align}
It is apparent that this cost function is designed to minimize the loss in rate \ie when a rate which is lower than the true rate is picked the packet transmission will be successful but there is an obvious loss in efficiency and this loss is the cost incurred. On the other hand, if a higher rate is picked then there is a packet loss and we lose the true rate that we could have got, entirely. This biases the predictor to pick lower values than the MAP predictor, thus leading to a lower packet loss.

 \section{Simulations, Results and Inference} \label{res}
Two cases of loading are considered i.e. a) Partial Loading,\footnote{For more details on partial loading refer to the Section \ref{symmo}} b) Full Loading. For both these cases, we use the MCS sequences over 5000 sub-frames obtained from the full System Simulator as discussed earlier, for 210 users. This results in 210 sequences - one for each user, of length 1000, since, CQI feedback happens only once in every 5 sub-frames as discussed.

 We also analyzed the MCS sequences generated for each UE in order to understand the behaviour of the sequences, in the case of partial and full loading. From the sequences $\x$ we generated an absolute difference sequence by computing $|\xud-\xun|$ for all $n$ and studied the statistics of this new sequence for all UEs. For each user this sequence can indicate the extent of variability of the MCS value at $n$ and $n+\delta$. It was found that $35\%$ of the users exhibited variations greater than 3 between adjacent values ($\xud=\xun\pm3$) for atleast 200 times in a 1000 length sequence for partial loading, while only $5\%$ of users under full loading had ($\xud=\xun\pm3$) for more than 50 times in a 1000 length sequence. For example, an MCS value of 15 could change to 12 or 18 before the next feedback, i.e., from a bits per symbol rate of 1.96 one will go down to 1.33. Similarly $20\%$ of the users had variations greater than 4 between adjacent values ($\xud=\xun\pm4$) for atleast 200 times in a 1000 length sequence for partial loading while there was not a single user with more than 25 such events in full loading. All of this points to a high degree of variability in the MCS sequence for partial loading. Hence outdated MCS seems to be a critical issue in partial loading.

For each user sequence ${\x_1.\x_2 ... \x_{1000}}$, the following prediction procedure is implemented on the system simulator
\begin{enumerate}
 \item  We build frequency trees upto depth $m$, which are updated as and when the sequence arrives. We choose $m=5$ since we are looking only at a sequence of length thousand \footnote{We have restricted the sequence length to 1000 due to a) the presence of UE sleep cycle and, b) assumption of stationarity of sequence may not hold over a long sequence length. }. This can be increased to $m=8$ or higher, if one has access to longer sequences.
 \item Then, using the frequency trees the probabilities $P(\xun|\x_{n-1}..\x_{n-k})$ are calculated as discussed earlier using  \eqref{eq:p1},\eqref{eq:p2} with $k=1 \hdots 4$. 
 \item $\Ipred(k)$ is then calculated online \ie as each value is received,  we use the probabilities obtained in Step 2 in  \eqref{eq:iprk}, to compute the empirical value of $\Ipred(k)$ using the probabilities and sequences seen so far.
   At time $n$ the sequence $\x_{n-1}..\x_{n-k}$ is used to calculate $P(\xun|\x_{n-1}..\x_{n-k})$ and these probabilities are used as follows to find the instantaneous predictive information of the sequence:
   \ifCLASSOPTIONtwocolumn
   \begin{align}
   \Ipred(k,n)=log(p)-\sum_{\xun=1}^p &P(\xun|\x_{n-1}..\x_{n-k})\nonumber\\&log(P(\xun|\x_{n-1}..\x_{n-k}))
  \end{align}
   \else
  \begin{align}
   \Ipred(k,n)=log(p)-\sum_{\xun=1}^p P(\xun|\x_{n-1}..\x_{n-k})log(P(\xun|\x_{n-1}..\x_{n-k}))
  \end{align}
  \fi
 This value of $\Ipred(k,n)$ is then empirically averaged over $n$, to get the current online estimate of $\Ipred(k)$ as follows:
 \begin{align*}
 \Ipred(k)=\frac{1}{n}\sum_{i=1}^n\Ipred(k,i)
 \end{align*}
 \item From the $\Ipred(k)$ obtained in Step 3, using  \eqref{eq:lk} which is the learning curve based stopping criterion, the value of $\kuopt$ is found for each user once the sequence length reaches 100, and this step is repeated once in every 100 values\footnote{The sequence should be of a sufficient length to get a reasonable average.} of the sequence i.e. $n=200,300$ and so on. It will take time to build a reasonably informative frequency tree for prediction. Hence, till the sequence length reaches 100 we do prediction using a simple Markov model i.e., we do not wait for a training period before starting prediction.
 \item Using $\kuopt$ as an upper bound on the model order, the optimal model order when the distribution is unknown $\kuoptt$, is found out using \eqref{eq:aiccmod}, \eqref{eq:aim} once the sequence length reaches 100 ,and this is also repeated once in every 100 values of the sequence.
 \item Then the tree is virtually truncated at depth $\kuoptt+1$. 
 \item This tree is used to find the probabilities $P(\xun|\x_{n-1}..\x_{n-\kuoptt})$ which are now used in the prediction algorithm.
 \item  These probabilities $P(\xun|\x_{n-1}..\x_{n-\kuoptt})$ obtained from Step 7) are used for prediction. We compare this with probabilities obtained from a virtually truncated tree of fixed depth $4$. The tree of fixed depth $4$ gives us the probabilities $P(\xun|\x_{n-1},\x_{n-2},\x_{n-3})$. The predictors using $P(\xun|\x_{n-1}..\x_{n-\kuoptt})$ and $P(\xun|\x_{n-1},\x_{n-2},\x_{n-3})$ are henceforth referred to as Variable Order (VO) predictors and Fixed Markov (FM) predictors respectively.

\end{enumerate}

  We use the probabilities computed using PPM with VO and FM in the MAP predictor in  \eqref{eq:map} and in the Bayesian Risk Mimimizer (BRM) presented in Section \ref{brm} in  \eqref{eq:cost} and compare the performance of the four schemes namely, FM-MAP, FM-BRM, VO-MAP and VO-BRM.  In \cite{marmon2007} nine techniques are proposed for prediction and out of those the median technique where the median of previous $n$ CQI values is taken, performs best for vehicular users. Since we have Doppler see Table \ref{table:bsm} and partial loading, we compare our schemes with the median technique in \cite{marmon2007}. A naive algorithm with no prediction \ie when the previous value is used as it is, is also compared with the above given techniques.

We compare the various schemes based on the following metrics:
\begin{itemize}
 \item Packet loss fraction ($P^u_{loss}$): We compute the packet loss fraction for each user and it is given by:
\begin{align}
 P^u_{loss}=\frac{\sum_{n=1}^{P} \i({\xunh>\xun})}{P}
\end{align}
where $P$ is the total number of packets transmitted. Packet loss occurs whenever the predicted MCS is greater than the actual MCS .
 \item Rate Efficiency Percentage($r^u_{eff}$): The rate obtained due to the a specific prediction scheme is compared with the rate obtained if there was ideal prediction and $r^u_{eff}$ for each user is : 
\begin{align}
r^u_{eff}=
\begin{cases}
 \dfrac{Rate_{currentscheme}}{Rate_{ideal}} , \:\:\:\:Rate_{currentscheme}\leq Rate_{ideal} \\
   0 , \:\:\:\: Rate_{currentscheme}>Rate_{ideal}  
\end{cases}
\end{align}

 \end{itemize}
It is well known that one can reduce packet loss by reducing the MCS and transmitting at increasingly conservative rates. However, our schemes reduce the packet loss and at the same time improve rate efficiency, since they exploit the fact that one can learn/predict current MCS value by analyzing the complexity of the MCS sequence. Moreover, since MCS sequences of different UEs have varying complexities, we use independent learning mechanisms for each UE. 

Since there are 210 users, for both partial and full loading, the empirical Cumulative Distribution Function (CDFs) are plotted for all the above mentioned metrics and these are discussed in detail.
%
\ifCLASSOPTIONtwocolumn
\begin{figure}
\centering
\subfloat[][ Partial Loading]{
 \includegraphics[width=0.9\columnwidth]{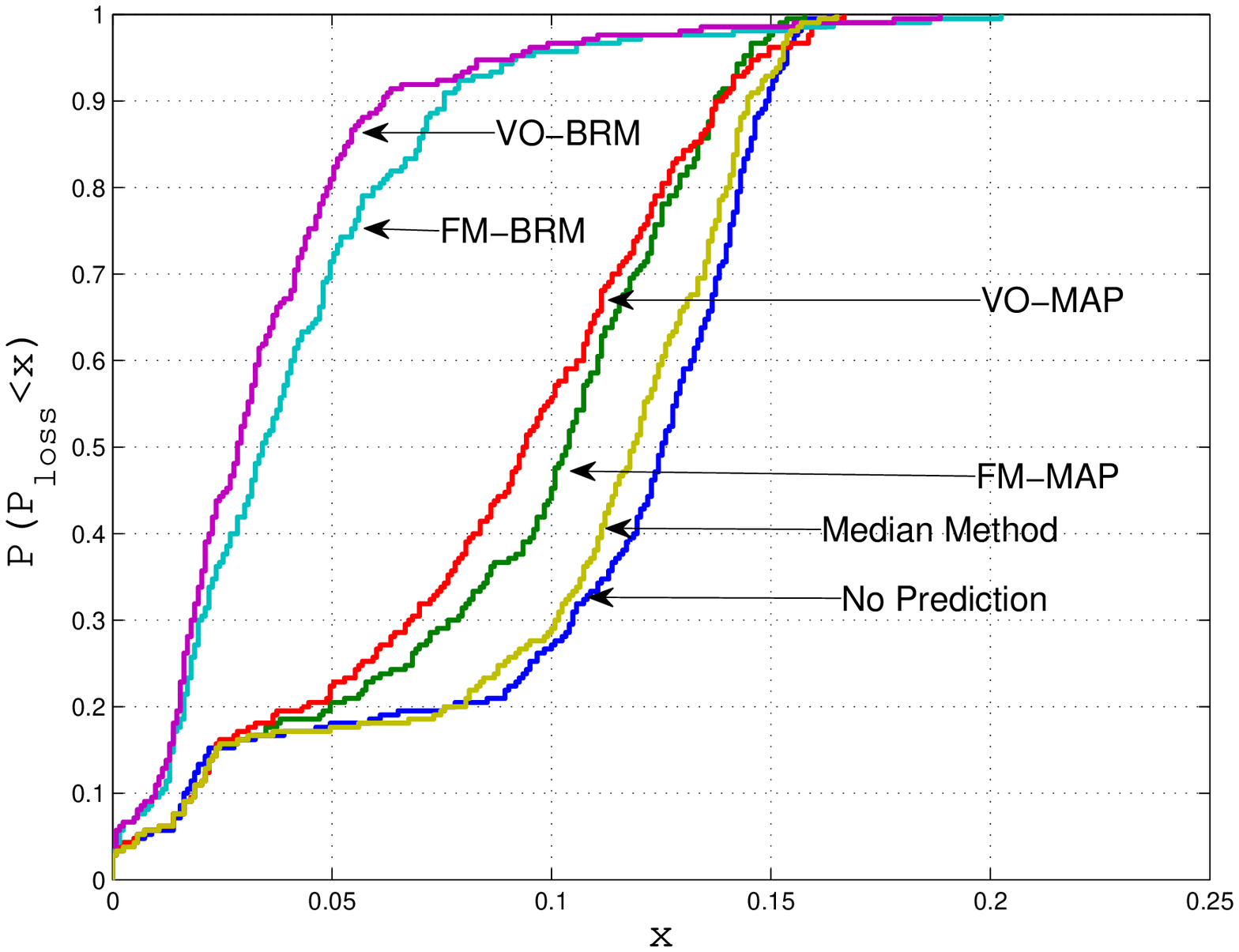}
\label{fig:harq}}
\qquad
\subfloat[][Full Loading]{
 \includegraphics[width=0.9\columnwidth]{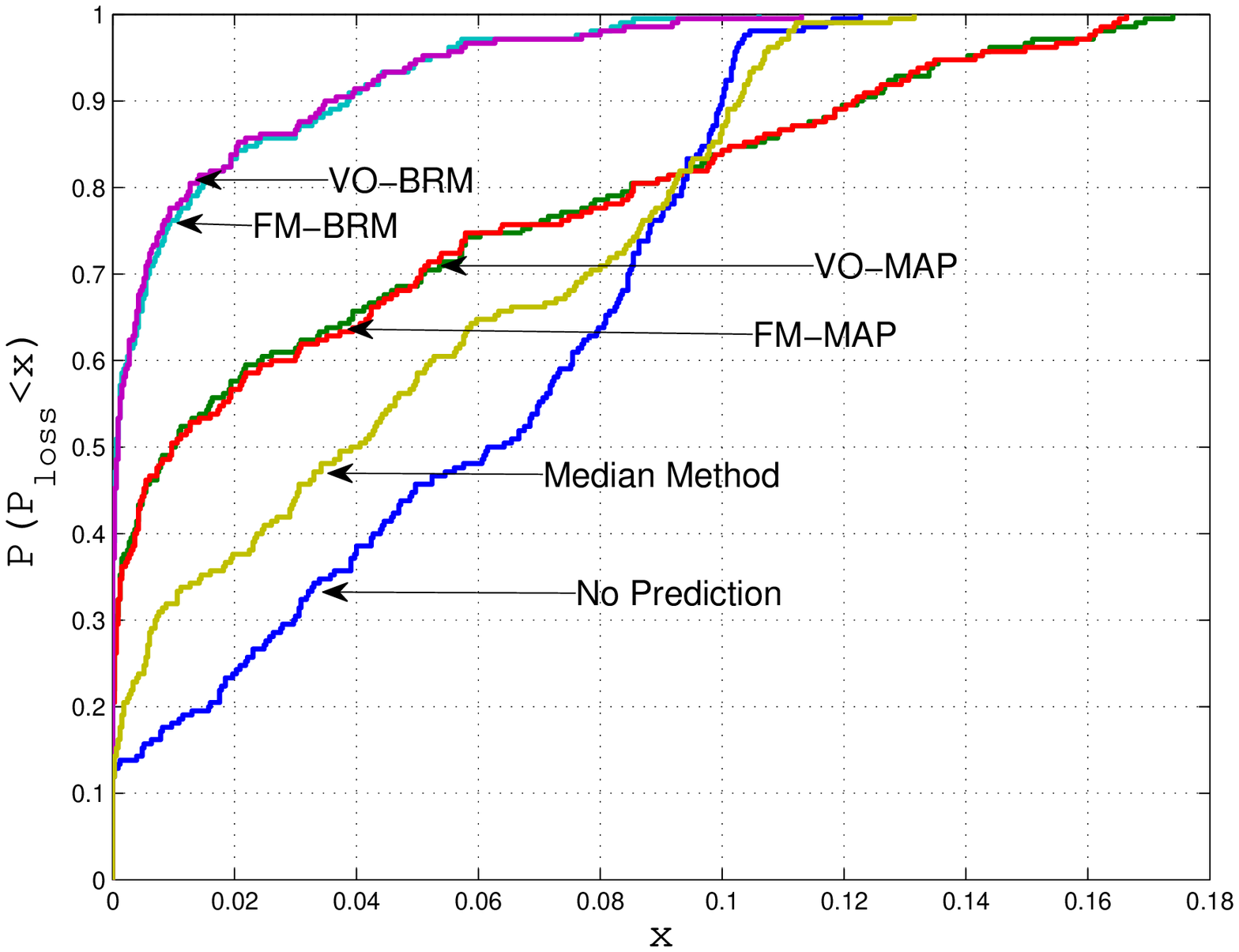}
\label{fig:harqinf}}
\caption{Packet Loss Fraction CDFs}
\label{fig:harqt}
\end{figure}
\else
\begin{figure}
\centering
\subfloat[][ Partial Loading]{
 \includegraphics[width=0.45\columnwidth]{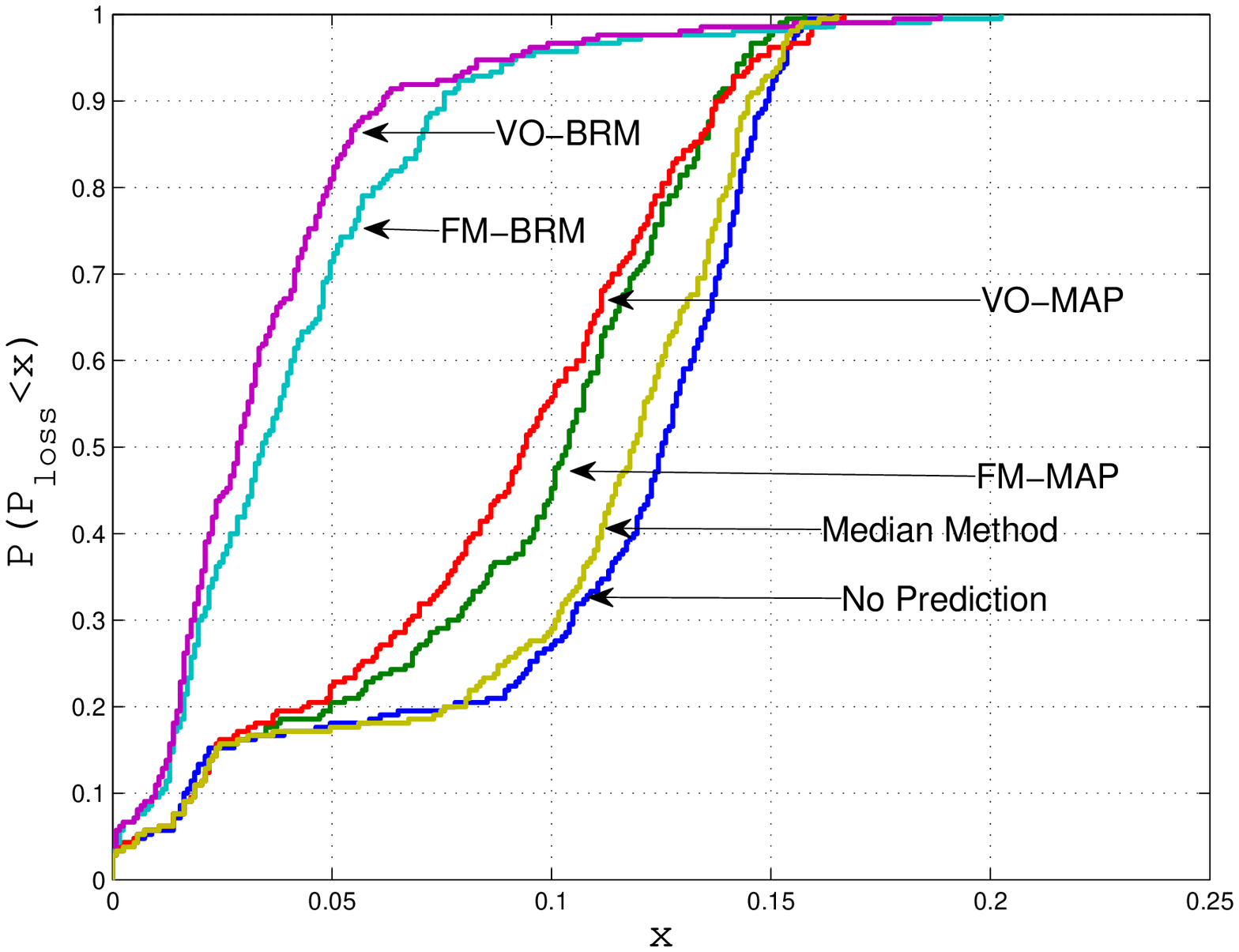}
\label{fig:harq}}
\qquad
\subfloat[][Full Loading]{
 \includegraphics[width=0.45\columnwidth]{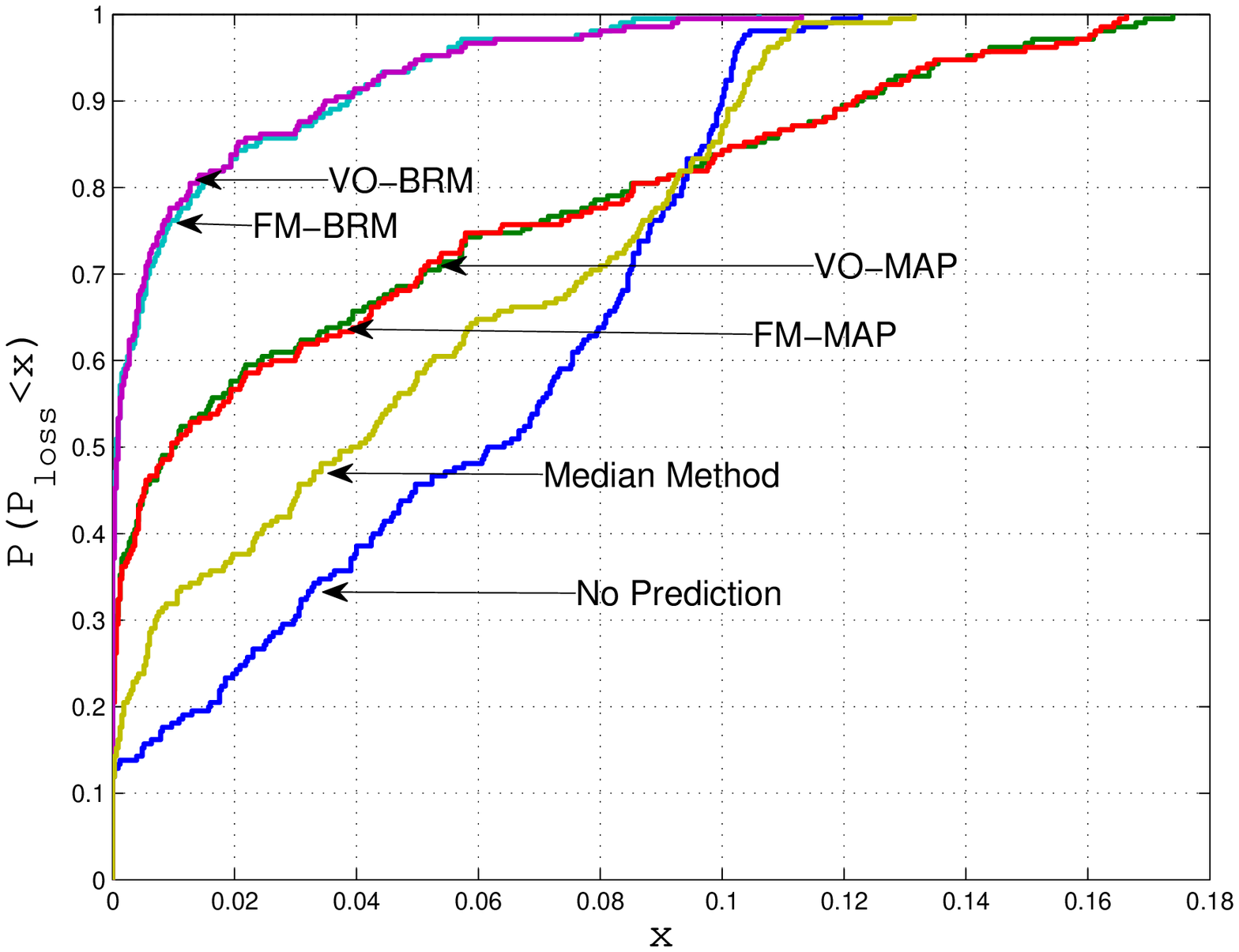}
\label{fig:harqinf}}
\caption{Packet Loss Fraction CDFs}
\label{fig:harqt}
\end{figure}
\fi
The packet loss fraction CDF under partial loading, is compared in \figref{fig:harq} and here it can be seen that the BRM predictors significantly outperform all other methods by having the lowest percentage of failed transmissions.When the VO-BRM method is used, $90\%$ of the users have less than $6.3\%$ packet loss, while when FM-BRM is used the corresponding packet loss is $7.6\%$. In comparison the VO-MAP, FM-MAP, Median and No Prediction have only $35\%$, $30\%$, $22\%$ and $20\%$ users with packet loss rate less than $7.6\%$. At the $50$-percentile point \footnote{corresponds to packet loss seen by at least $50\%$ of the users} in the packet loss distribution, VO-BRM at $2.8\%$ packet loss, outperforms the FM-BRM by $20\%$  and the VO-MAP and FM-MAP schemes by more than $200\%$ and $250\%$ respectively,  median scheme proposed in \cite{marmon2007} by $400\%$ and the no prediction scheme by nearly $450\%$.  This gain in packet loss performance is achieved with no loss in rate. 
\ifCLASSOPTIONtwocolumn

 \begin{figure}
\centering
\subfloat[][Partial Loading]{
 \includegraphics[width=0.9\columnwidth]{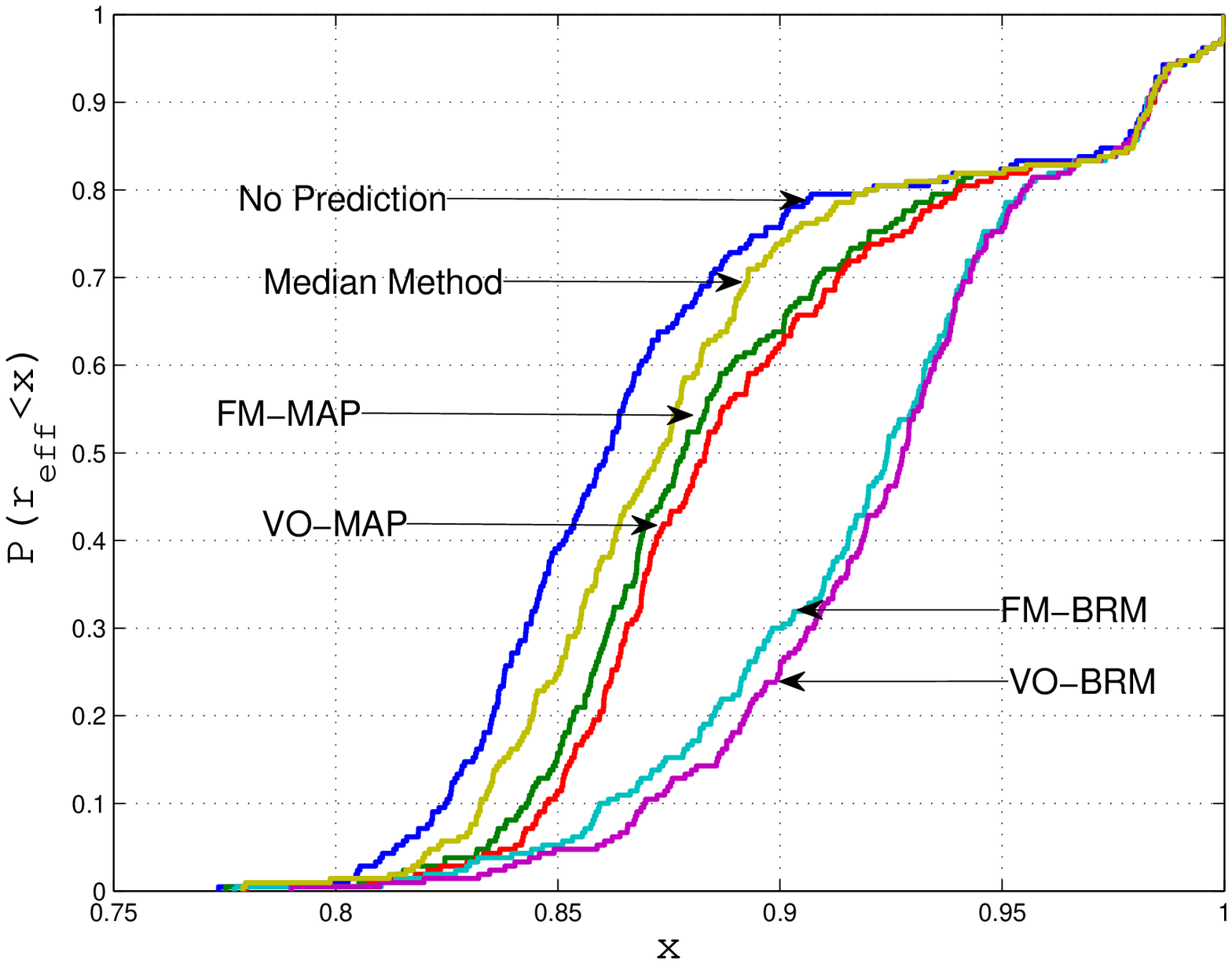}
\label{fig:reff}}
\qquad
\subfloat[][Full Loading]{
 \includegraphics[width=0.9\columnwidth]{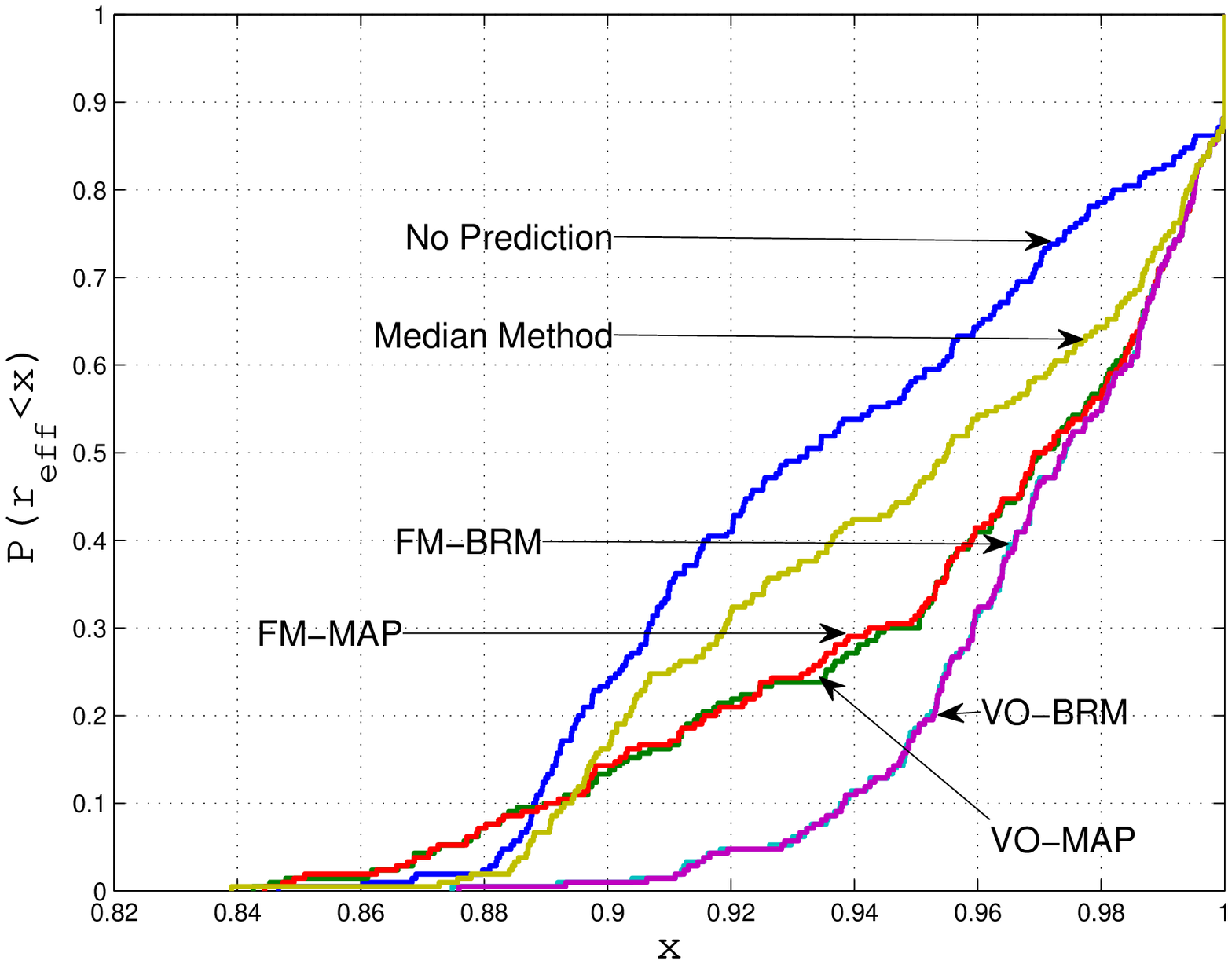}
\label{fig:reffinf}}
\caption{Rate Efficiency CDFs}
\label{fig:refft}
\end{figure}

\else
 \begin{figure}
\centering
\subfloat[][Partial Loading]{
 \includegraphics[width=0.45\columnwidth]{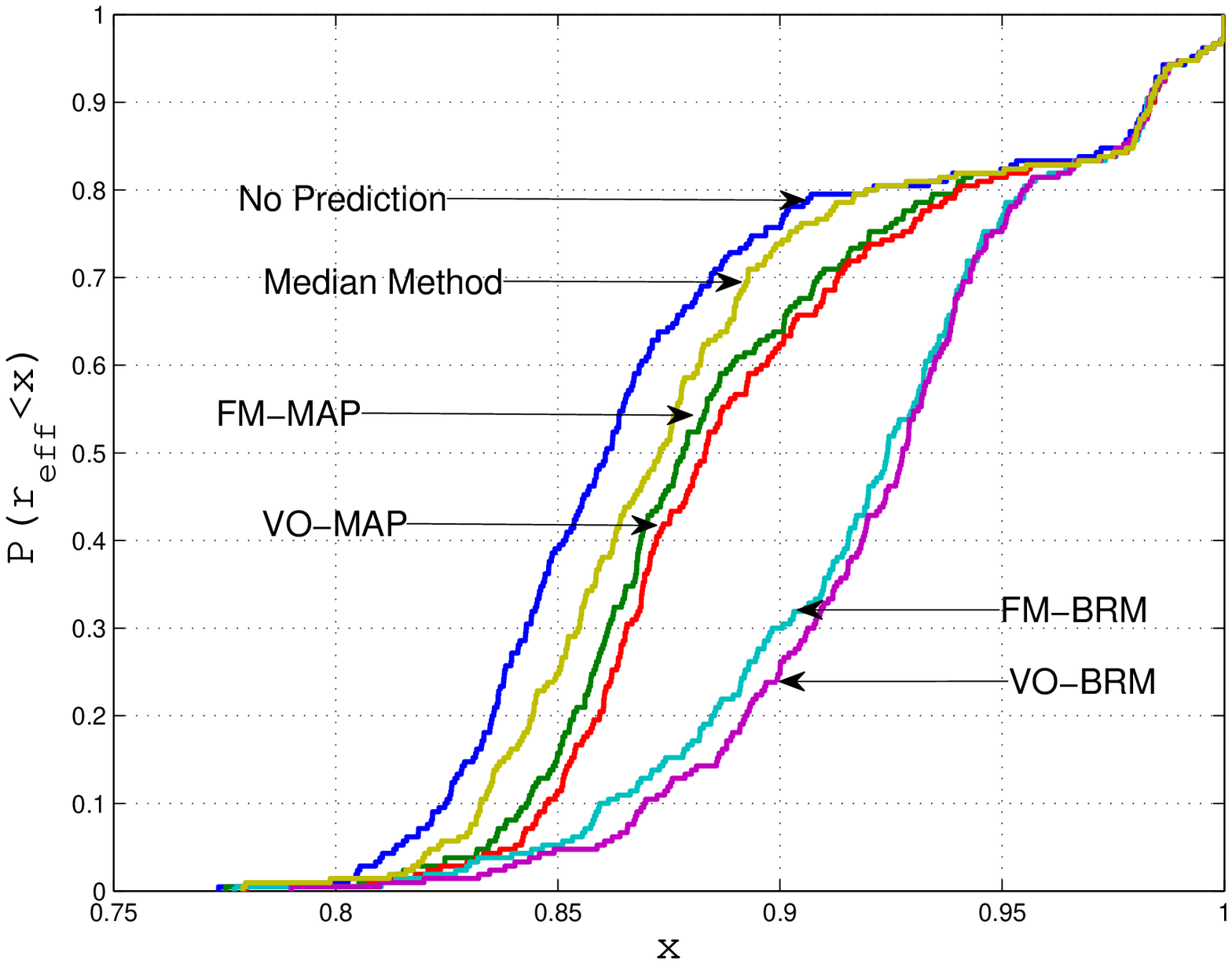}
\label{fig:reff}}
\qquad
\subfloat[][Full Loading]{
 \includegraphics[width=0.45\columnwidth]{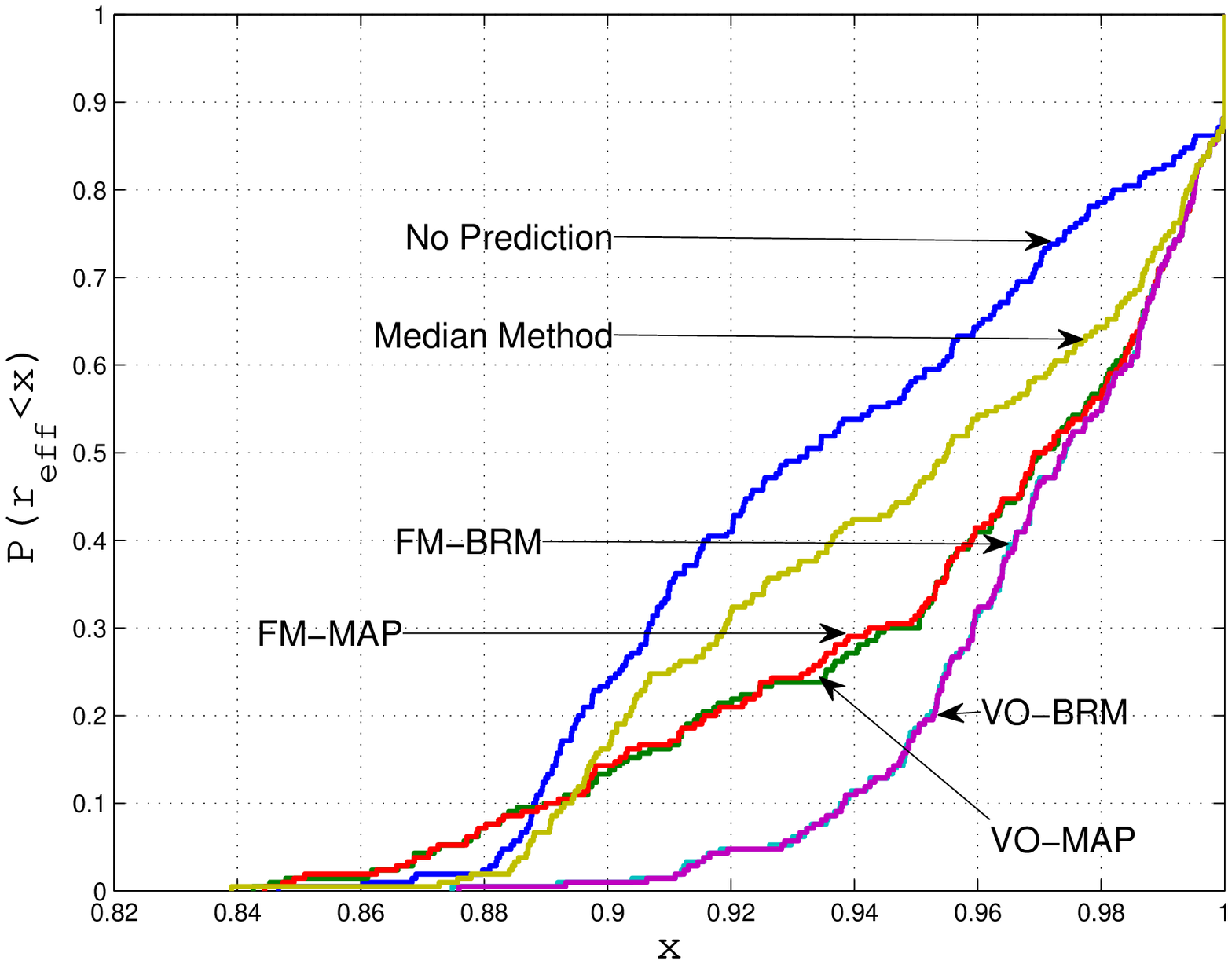}
\label{fig:reffinf}}
\caption{Rate Efficiency CDFs}
\label{fig:refft}
\end{figure}
\fi

 The rate efficiency CDF under partial loading is compared in \figref{fig:reff} and here again it can be seen that the BRM outperforms all other methods by having the highest rate efficiency. Here, VO-BRM has $76\%$ users achieving a rate efficiency of $90\%$ or higher, while FM-BRM had only $69\%$ users with this criteria.  This implies that while 160 users achieve a high rate efficiency using VO-BRM, only 146 users achieve the same using FM-BRM. The corresponding percentage of users with that rate efficiency were $38\%$, $35\%$,$26\%$ and $23\%$ for VO-MAP, FM-MAP, median technique and scheme without prediction respectively. 

When we look at full loading performance graphs in  \figref{fig:harqinf} and \figref{fig:reffinf} we can see that the trends of MAP versus BRM are similar \ie BRM is way better than MAP in packet loss percentage and in rate efficiency. There is a cross-over between the MAP and no prediction CDFs in packet loss percentage as seen in \figref{fig:harqinf}. This is because of the behavior of the MAP predictor where all errors are treated equal. Especially, when MAP predicts an MCS that is higher than the previous fed-back value and it is also higher than the true value, a packet loss occurs. Therefore, for some users the no prediction scheme performs better than MAP prediction. This effect is seen in the full loading scenario because, the MCS variation itself is likely to be more gradual and even without prediction, sometimes the fed-back MCS works better than a predicted MCS. However, on an average the MAP is better than not predicting and BRM is far better than both.
 However, when one compares FM to VO, it can be seen that, there is little to choose between them across all the performance metrics considered under full loading. This implies that partial loading requires us to adapt the model order, while, full loading performance may not require us to adapt the model order. Since all practical systems see partial loading, either due to traffic or due to sub-frame blanking, VO based methods are required to fully exploit the advantages of rate adaptation. 
%


%
%
%

\section{Conclusions}
The effect of outdated MCS in the presence of partial loading was investigated. Discrete sequence prediction algorithms such as PPM were proposed for MCS prediction. The optimal tree depth that one needs to traverse for prediction using PPM was cast as a model order problem. Techniques such as MDL, AIC and Corrected AIC were proposed to estimate the model order of the sequence for each user with the sequence complexity analysis providing an upper bound on the model order. Finally, the MAP and Bayesian Risk minimization based rate predictors were proposed and implemented for MCS prediction. Simulation results indicates that, using different model order for different users, gives substantial system level gains over assuming a fixed model order for all users. The gains due to adapting the model order, were found to be substantial in partially loaded systems. Furthermore, the proposed Bayesian Risk Minimization predictor, significantly outperforms the MAP based predictor.

 \bibliography{learning_REF.bib}
\bibliographystyle{IEEEtran}
\end{document}

%% file: notation.tex
\def\figref#1{Fig.\,\ref{#1}}%
\def\ie{{\em i.e., }}

\def\S{\mathbf{S}}

\def\i{\mathbf{1}}

\def\S{\mathtt{S}}
\def\one{\mathbf{1}}

\def\sinr{\mathtt{SINR}}

\def\xud{{X_{n+\delta}^u}}
\def\xudh{{\hat{X}_{n+1}^u}}
\def\xun{{X_{n}^u}}
\def\xunh{{\hat{X_{n}^u}}}

\def\x{{X_{ }^u}}
\def\ku{k^u}

\def\H{\mathcal{H}}
\def\kuopt{k^u_{opt}}
\def\km{k^{max}}

\def\kuoptt{\tilde{k}^u_{opt}}
\def\niu{n_i^u}
\def\Ipred{I_{pred}}
\def\thet{\boldsymbol {\theta_i}}

\def\wl{W_L}
\def\wlm{W_{L_{max}}}
  